%% file: review.tex
\documentclass[11pt]{article}
\pdfoutput=1
\usepackage{microtype}
\usepackage{amsmath,amssymb,amsthm} 
\usepackage{bbm}
\usepackage{stmaryrd}
\usepackage{paralist}
\usepackage{hyperref}
\usepackage{graphicx}
\usepackage{tikz}
\usepackage{pgfplots}
\usepackage{color}
\usepackage{commath, mathtools}

\setdefaultenum{(i)}{}{}{}
\usepackage[margin=1in]{geometry} 
\usepackage[color=yellow]{todonotes}

\usetikzlibrary{matrix, calc}

\numberwithin{equation}{section}

\numberwithin{equation}{section}

\renewenvironment{thebibliography}[1]{%
\begin{oldthebibliography}{#1}%
\setlength{\baselineskip}{.8em}
\linespread{.8}
\small
\setlength{\parskip}{0ex}%
\setlength{\itemsep}{.1em}%
}%
{%
\end{oldthebibliography}%
}

\theoremstyle{plain}
\newtheorem{theorem}{Theorem}[section]

\newtheorem{definition}[theorem]{Definition}

\newcommand{\E}{\mathbb{E}}
\newcommand{\R}{\mathbb{R}}

\DeclareMathOperator{\Tr}{Tr}
\DeclareMathOperator{\diag}{diag}
\DeclareMathOperator*{\argmin}{arg\,min}

\begin{document}

\title{A Primer on Portfolio Choice with Small Transaction Costs
\footnote{The second and third authors were partly supported by the ETH Foundation, the Swiss Finance Institute, and Swiss National
Foundation grant SNF 200021\_153555. Parts of this paper were written while the first author was visiting the Forschungsinstitut Mathematik at ETH Z\"urich.}}
\date{May 23, 2017}
\author{Johannes Muhle-Karbe\thanks{University of Michigan, Department of Mathematics, 530 Church Street, Ann Arbor, MI 48109, USA, email \texttt{ johanmk@umich.edu}.} \and Max Reppen\thanks{ETH Z\"urich, Department of Mathematics, R\"amistrasse 101, 8092 Z\"urich, Switzerland, email \texttt{max.reppen@math.ethz.ch}.} \and H. Mete Soner\thanks{ETH Z\"urich, Department of Mathematics, R\"amistrasse 101, 8092 Z\"urich, Switzerland, email \texttt{mete.soner@math.ethz.ch}.}
}
\maketitle

\begin{abstract}
\noindent This survey is an introduction to asymptotic methods for portfolio-choice problems with small transaction costs. We outline how to derive the corresponding dynamic programming equations and simplify them in the small-cost limit. This allows to obtain explicit solutions in a wide range of settings, which we illustrate for a model with mean-reverting expected returns and proportional transaction costs. For even more complex models, we present a policy iteration scheme that allows to compute the solution numerically.
\end{abstract}

\input{introduction.tex}

\input{setting.tex}

\input{dynamic_programming.tex}

\input{homogenization.tex}

\input{examples.tex}

\input{extensions.tex}

\input{numerics.tex}

\input{appendix.tex}

\bibliographystyle{abbrv}
\bibliography{references}

\end{document}

%% file: introduction.tex
\section{Introduction}

Starting with the work of Constantinides and Magill \cite{constantinides.magill.76}, Constantinides  \cite{constantinides.86}, Dumas and Luciano \cite{dumas.luciano.91}, Davis and Norman \cite{davis.norman.90}, and Shreve and Soner \cite{shreve.soner.94}, models with \emph{transaction costs} have been subjected to intensive research. For example, much effort has been devoted to understanding liquidity premia in asset pricing \cite{constantinides.86,jang.al.07,lynch.tan.11,dai.al.16} or how transaction costs shape the trading volume in financial markets \cite{scheinkman.xiong.03,lo.al.04,gerhold.al.14}. On a more practical level, transaction costs play a crucial role in the design and implementation of trading strategies in the asset management industry, cf., e.g., \cite{grinold.06,martin.schoeneborn.11,bouchaud.al.12,martin.12}.

However, the quantitative analysis of models with trading costs is quite difficult. Unlike in frictionless models, the position in each asset becomes a state variable because it can no longer be adjusted immediately and for free. As a consequence, explicit solutions are no longer available even in the simplest models with constant market and preference parameters and an infinite planning horizon \cite{dumas.luciano.91,davis.norman.90,taksar.al.88}. This is only compounded in more complex models with random and time-varying investment opportunities. However, transaction costs become crucially important precisely in such settings where, for example, prices exhibit momentum or mean reversion \cite{martin.schoeneborn.11,martin.12,garleanu.pedersen.13,garleanu.pedersen.16,dufresne.al.12,bouchaud.al.12}, switch between different regimes \cite{jang.al.07}, or investors are exposed to idiosyncratic endowment shocks \cite{lo.al.04,lynch.tan.11}. Then, investors can no longer ``accommodate large transaction costs by drastically reducing the frequency and volume of trade'' \cite{constantinides.86} as in simple models where portfolio rebalancing is the only motive to trade. Instead, striking the right balance between adjustments to optimize performance and the induced implementation costs then becomes a central issue.

To obtain tractable results in complex models with transaction costs, it is often very useful to take an asymptotic perspective and view them as small perturbations of a frictionless benchmark model. The goal then is to obtain \emph{explicit asymptotic formulas} for optimal trading policies and the associated welfare effect of small transaction costs. Results of this kind were first obtained in simple concrete models that can be solved explicitly in the frictionless case \cite{dixit.91,shreve.soner.94,whalley.wilmott.97,janecek.shreve.04,lo.al.04,bichuch2013utility,bichuch.12,gerhold.al.14}. In the last couple of years, there has been a lot of progress in extending these sensitivity analyses to much more general settings \cite{bichuch.14, bouchard2014hedging, soner.touzi.13,possamai.al.15,altarovici.al.15,moreau.al.15,martin.12,kallsen.muhlekarbe.15a,kallsen.muhlekarbe.15,rosenbaum.tankov.14,cai.al.15,cai.al.16,ahrens.15,feodoria.16}. These results are obtained using a range of different methods, ranging from analytic studies of the dynamic programming equation \cite{soner.touzi.13,possamai.al.15, altarovici.al.15,moreau.al.15,martin.12} to convex duality arguments \cite{ahrens.15}, and weak-convergence techniques \cite{cai.al.15,cai.al.16}.

In this survey, we review the \emph{homogenization approach} put forward in \cite{soner.touzi.13} for models with proportional transaction costs. This approach based on partial differential equations is very flexible and readily adapts to many variations of the model, e.g., different cost structures \cite{altarovici.al.15,altarovici.al.16,moreau.al.15} or preferences \cite{bouchard2014hedging,melnyk.al.16}. Hence, similarly to the classical dynamic approach to frictionless control problems (see, e.g., \cite{fleming.soner.06} for an overview), this method has the potential to be a key tool for the analysis of a wide range of complex models.\footnote{``Martingale methods'' based on the duality theory introduced by \cite{cvitanic.karatzas.96} allow to go beyond the Markovian paradigm \cite{kallsen.muhlekarbe.15a,kallsen.muhlekarbe.15,ahrens.15}. However, they are more difficult to adapt to new models and not applicable if the problem at hand is not convex, e.g., due to the presence of fixed trading costs.} The present review is written as a user's guide for the application of this method. We explain in detail both the basic underlying ideas and each step of their application to a concrete problem. The goal is to provide a blueprint that will allow the readers to apply the approach to a wide range of related problems. 

To illustrate the effects of transaction costs on active investment in a concrete example, we also provide a detailed discussion of a model with mean-reverting returns. Starting from the frictionless solution of~\cite{kim.omberg.96}, we show how both the leading-order optimal trading policy and the corresponding performance loss due to the trading friction can be computed explicitly. We show that -- like for models with constant investment opportunities \cite{constantinides.86} -- this leads to wide no-trade regions and a severe reduction of trading volume. However, the corresponding welfare losses are greatly amplified -- the opportunity cost of not being able to ``time the market'' is substantial. This underlines the importance of correcting the performance of active investment strategies for trading costs (compare \cite{bajgrowicz.scaillet.12} and the references therein), for which the tractable asymptotic formulas reviewed here provide a convenient analytical tool.

This review is organized as follows. We first introduce our continuous-time model without and with transaction costs in Section~\ref{sec:model}. Then, we derive the corresponding frictionless and frictional dynamic programming equations. These partial differential equations for the value functions of the problems at hand are the starting point of the subsequent analysis. Afterwards, in Section~\ref{sec:homogenization}, we outline the homogenization approach, and discuss in detail how to apply it to models with proportional transaction costs. This method allows to reduce the complexity of the problem at hand by reducing the number of state variables, simplifying the underlying state dynamics, and postponing finite time horizons to infinity. In Section \ref{sec:examples}, we explain how this allows to obtain explicit solutions in the model of Kim and Omberg \cite{kim.omberg.96}, where asset prices exhibit momentum. Subsequently, Section~\ref{s.extension} provides references to several extensions of the homogenization results to more general asset dynamics, preferences, and cost structures. Finally, in Section \ref{sec:policy_iteration}, we discuss a numerical scheme that allows computation of the solution to the simpler ``homogenized'' problem numerically using a policy iteration algorithm. To focus on the main ideas and computational issues, mathematical formalism is treated liberally throughout this survey. Rigorous verification theorems for the results presented here can be found in \cite{soner.touzi.13,rosenbaum.tankov.14,moreau.al.15,altarovici.al.15,possamai.al.15,ahrens.15,feodoria.16,cai.al.15,cai.al.16,melnyk.seifried.15}.

\paragraph{Notation} We write $D_x\varphi((t,x,y,f), D_{xf}\varphi((t,x,y,f)$, etc.\ for the partial derivatives of a multivariate function $\varphi(t,x,y,f)$. When there is no confusion, we also use the 
more compact notation $\varphi_x$, $\varphi_{x f}$, etc. As is customary in asymptotic analysis, $\mathcal{O}(\delta)$ denotes any function satisfying $\left| \mathcal{O}(\delta)\right| \le C \delta$ for a constant $C>0$ and all $\delta \in [0,1]$.
For any integrable random variable $\xi$ and a time point $t \ge 0$, $E_t[\xi]$ denotes the expectation of $\xi$ conditional on the information up to time $t$.

%% file: setting.tex
\section{Model}\label{sec:model}

\subsection{Financial Market}

We consider a financial market with one safe and one risky asset with dynamics modulated by a general factor process. More precisely, the safe asset follows
\[ \frac{\dif{B_s}}{B_s}=r(F_s) {\dif{s}}, \]
and the risky dynamics are
\begin{equation}\label{eq:risky}
    \frac{\dif{S_s}}{S_s}=(r(F_s)+\mu_S(F_s)) {\dif{s}}+\sigma_S(F_s) \dif{W^S_s}.
\end{equation}
Here, $(W^S_s)_{s \in [0,T]}$ is a standard Brownian motion; the \emph{safe rate} $r(f)$, the \emph{expected excess return} $\mu(f)$, and the \emph{volatility} $\sigma(f)$ are sufficiently smooth deterministic functions of the factor process $(F_s)_{s\in [0,T]}$. The latter follows an autonomous diffusion:
\begin{equation}\label{eq:facdyn}
\dif{F}_s=\mu_F(F_s)\dif{s}+\sigma_F(F_s)\dif{W^F_s}.
\end{equation}
Here, $(W^F_s)_{s \in [0,T]}$ is another standard Brownian motion that has constant correlation $\rho \in [-1,1]$ with the Brownian motion $(W^S_s)_{s \in [0,T]}$ driving the risky returns. Both $\mu_F(f)$ and $\sigma_F(f)$ are sufficiently regular deterministic functions.

\paragraph{Examples.} All tractable models from the literature fit into this framework. Examples are:
\begin{enumerate}
\item
\label{ex.bs}
\emph{Black--Scholes Model}: the standard example for the asset dynamics is the Black--Scholes model, where the safe rate, the expected risky return, and the volatility are all constants: $r(f)\equiv r$,
$\mu_S(f)\equiv \mu_S$, and $\sigma_S(f)\equiv \sigma_S$.  
\item
\label{ex.ko}
\emph{Kim--Omberg Model}: to study the effects of transaction costs in a model where investment opportunities vary randomly over time, we consider the model of Kim and Omberg \cite{kim.omberg.96}.\footnote{This is a standard model for the ``predictability of asset returns'', which has been discussed extensively in the empirical literature \cite{welch.goyal.08,cochrane.08}. The importance of transaction costs in such environments that require to ``time the market'' is evident and discussed in \cite{bouchaud.al.12,martin.schoeneborn.11,martin.12,garleanu.pedersen.13,moreau.al.15,garleanu.pedersen.16,dufresne.al.12}, for example.} This means the safe rate and volatility remain constant ($r(f)\equiv  r$, $\sigma_S(f)\equiv  \sigma_S$), but the expected excess returns follow an Ornstein--Uhlenbeck process:\footnote{The framework in \cite{kim.omberg.96} also allow other choices of $\mu^S$ and $\sigma^S$ for which the Sharpe ratio $\mu^S/\sigma^S$ remains an Ornstein--Uhlenbeck process, because all of these models span the same \emph{frictionless} payoff spaces. As this invariance breaks down with transaction costs, we focus on the present specification here.}
 $\mu_S(f)=f$ and
 \begin{equation}\label{eq:oudyn}
 \dif{F}_s=\kappa (\bar{F}-F_s){\rm{d}}s+\sigma_F \dif{W}^F_s,
 \end{equation}
for constants $\kappa$, $\bar{F}$ and $\sigma_F$ describing the mean-reversion speed, the mean-reversion level, and the volatility of the expected excess return. 
\item
\label{ex.heston}
\emph{Heston--type models}: in another widely used class of models, the volatility is assumed to be a mean-reverting process. For example, Heston \cite{heston1993closed} proposes a constant interest rate $r$ and excess return $\mu_S$, as well as a stochastic volatility $\sigma_S(f)= \sqrt{f}$ where the factor $F$ is a square root process.  Liu \cite{liu.07} instead sets $\mu_S(f)= \alpha f$ for some constant $\alpha$, retaining the other specifications of Heston. Chacko and Viceira \cite{chacko.viceira.05} keep Heston's constant $r$ and $\mu_S$, but their volatility is $\sigma_S(f)=\sqrt{1/f}$.
\end{enumerate}

\subsection{Trading and Optimization}

We now turn to trading and optimization in the above financial market. For concreteness, we focus on a specific portfolio choice problem where consumption only takes place at the terminal time. Extensions to more general settings do not pose any essential difficulties and are discussed in Section~\ref{s.extension}. We first briefly recall the frictionless case and then turn to models with transaction costs.

\paragraph{Frictionless Case}
\label{ss.frictionless}
Starting with an initial endowment of $x_0$ dollars in the safe account and a risky position worth $y_0$ dollars, an agent can trade the safe and the risky asset continuously on $[0,T]$. Without trading costs, positions can be changed freely over time, so that the amount of money $Y_t$ invested in the risky asset is therefore a suitable control variable. The wealth dynamics generated by such a strategy is simply obtained by weighting the safe and risky returns according to the corresponding investments:\footnote{The superscripts in our notation refer to the initial conditions $Z^{t,z,f}_t=z$ and $F^{t,f}_t=f$.}
\begin{equation}\label{eq:state}
\begin{split}
    \dif{Z}^{t,z,f}_s &= Y_s \left[(r(F^{t,f}_s)+\mu_S(F^{t,f}_s)){\rm{d}}s+\sigma_S(F^{t,f}_s)\dif{W}^S_s\right]
+(Z^{Y,x_0+y_0}_s-Y_s)r(F^{t,f}_s){\rm{d}}s, \quad s \geq t,\\
 F^{t,f}_t&=f, \quad Z^{t,z,f}_t=z=x_0+y_0.
\end{split}
\end{equation}

Without trading costs, the wealth $z$ is the only state variable we need to keep track of, apart from to the factor $f$. In contrast, the decomposition of $z$ into the risky position $y$ and the safe position $x=z-y$ is irrelevant because it can be changed instantly and without cost, by updating the control. If agents choose their trading strategies to maximize expected utility from terminal wealth at time $T$ for some utility function $U$, we therefore expect the \emph{value function} to be a deterministic function of the current time $t$, the current wealth $z$, and the current value $f$ of the factor only:
\begin{equation}\label{eq:vf1}
    v(t,z,f)=\,\,\sup_{\mathclap{(Y_s)_{s \in [t,T]}}}\,\,E_t\left[ U\left(Z^{t,z,f}_T\right)\right].
\end{equation}

\paragraph{Proportional Transaction Costs}
\label{ss.proportional}
Now, suppose that trades incur a cost $\lambda$ proportional to the value traded. Then, the decomposition of the total wealth evidently matters, because this ratio can no longer be adjusted for free. As a consequence, we need to keep track of the evolution of the safe and risky positions separately. These quantities now both become state variables that can only be adjusted gradually. To wit, agents now choose nondecreasing adapted process $L$ and $M$ that describe the cumulative transfers from the safe to the risky account and vice versa. The corresponding dynamics of the safe account are\footnote{This means that transaction costs are always deducted from the safe account, both for purchases and sales of the risky asset.}
\begin{equation}\label{e.cash}
\begin{split}
\dif{X}^{t,x,y,f}_s &=r(F^{t,f}_s)X^{t,x,y,f}_s {\rm{d}}s -(1+\lambda)\dif{L}_s+(1-\lambda)\dif{M}_s, \quad s \geq t,\\
X^{t,x,y,f}_t &=x.
\end{split}
\end{equation}
The dynamics of the risky account read as follows:
\begin{equation}\label{e.risky}
\begin{split}
    \dif{Y}^{t,x,y,f}_s &=Y^{t,x,y,f}_s
    \left[(r(F^{t,f}_s)+\mu_S(F^{t,f}_s)){\rm{d}}s +\sigma_S(F^{t,f}_s)\dif{W}^S_s\right] 
    +\dif{L}_s-\dif{M}_s, \quad s \geq t,\\
Y^{t,x,y,f}_t&=y.
\end{split}
\end{equation}
If agents maximize expected utility form terminal paper wealth,\footnote{One could also consider expected utility from \emph{liquidation wealth} $X^{t,x,y,f}_T+(1-\lambda)Y^{t,x,y,f}_T$, but this does not affect the asymptotic results at the leading order.} the corresponding frictional value function will in turn depend on the current values of the safe \emph{and} risky account in addition to time and the current value of the factor process:
\begin{equation}\label{eq:vf2}
    v^\lambda(t,x,y,f)=\quad \sup_{\mathclap{(L_s,M_s)_{s \in [t,T]}}}\quad E_t\left[ U\left(X^{t,x,y,f}_T+Y^{t,x,y,f}_T\right)\right].
\end{equation}

\paragraph{Fixed and Proportional Transaction Costs}
\label{ss.fixed}

In this survey, we mostly focus on the above model with proportional transaction costs in order to describe the main ideas of the homogenization approach for small transaction costs most clearly. However, the methods outlined here readily adapt to more general settings \cite{moreau.al.15,altarovici.al.15,altarovici.al.16}. For example, in a model with proportional costs $\lambda_p$ and additional fixed costs $\lambda_f$ per trade, the cash dynamics \eqref{e.cash} change to
\[  X^{t,x,y,f}_s =x + \int_t^s r(F^{t,f}_\tau)X^{t,x,y,f}_\tau \dif{\tau}- L_s +M_s
    -\lambda_p(L_s+M_s) 
    - \lambda_f J_s(L,M), \quad s \geq t, \]
where $J_s(L,M)$
is the total number of jumps of $L$ and
$M$ up to time $s$.
Since one pays a fixed, non-zero amount for each jump,
$J_s(L,M)$ must be finite for any admissible strategy, unlike for proportional costs. Nevertheless, the value function depends on the same variables as in~\eqref{eq:vf2}.

%% file: dynamic_programming.tex
\section{Dynamic Programming}

 The key concept for studying the value functions \eqref{eq:vf1}, \eqref{eq:vf2} and the corresponding optimal trading strategies is to describe them in terms of a partial differential equation derived from the \emph{dynamic programming principle} of stochastic optimal control. Loosely speaking, the latter states that if we have already determined the optimal policy on $[t+\dif{t},T]$, then fixing this policy and optimizing over the choice at time $t$ leads to the same solution as optimizing over the entire interval $[t,T]$. In discrete time, this means that the optimal policy can be computed by backward induction; in the continuous-time limit, a partial differential equation is obtained.\footnote{Here, we only provide a heuristic derivation of the dynamic programming equations. For a mathematically
 rigorous treatment we refer the reader to the monograph \cite{fleming.soner.06}; more recent results can be found in \cite{bouchard.touzi.11,bouchard.nutz.12,altarovici.al.15,altarovici.al.16,elkaroui.tan.13}.}  
 \subsection{Frictionless Case}
 
 Let us illustrate this idea by first briefly recalling the frictionless case. Then, the dynamic programming principle suggests that
 $$v(t,z,f) = \sup_{Y_t} E_t\left[v\left(t+\dif{t},z+\dif{Z}^{t,z,f}_t,f+\dif{F}^{t,f}_t\right)\right].$$
 Now, apply It\^o's formula, insert the state dynamics (\ref{eq:risky}--\ref{eq:facdyn}), and cancel the stochastic integrals (because they are martingales and therefore have zero expectation if the integrands are sufficiently integrable). Finally, divide by $\dif{t}$ and send $\dif{t}$ to zero. Dropping the arguments to ease notation, this leads to the following equation:
\begin{equation}\label{eq:dpe}
\begin{split}
    0 &= v_t + \mu_F v_f +\frac12 \sigma_F^2 v_{ff} + \sup_{Y_t} \Big\{ \left(Y_t(r+\mu_S)+(z-Y_t)r\right) v_z +\frac12 Y_t^2 \sigma_S^2 v_{zz} + Y_t \rho \sigma_S\sigma_F v_{zf} \Big\}.
 \end{split}
\end{equation}
The above equation incorporates the influence of the momentary choice for the evolution of the system into the value function. Hence, the problem at time $t$ is reduced to a simple optimization over instantaneous controls. In this particular case, we obtain a pointwise quadratic problem in $Y_t$, whose solution is given by 
\begin{equation}\label{eq:feedback}
    Y_t =\theta(t,z,f):=-\frac{\mu_S(f)}{\sigma_S^2(f)} \frac{v_z(t,z,f)}{v_{zz}(t,z,f)}-\frac{\rho\sigma_F(f)}{\sigma_S(f)}\frac{v_{zf}(t,z,f)}{v_{zz}(t,z,f)}.
\end{equation}
Plugging this expression back into the dynamic programming equation \eqref{eq:dpe} yields
\begin{equation}
    \label{eqn:frictionlessoperator}
    \mathcal{A} v := v_t + \mu_F v_f + \frac{1}{2} \sigma_F^2 v_{ff} + r z v_z + \mu_S \theta v_z + \frac{1}{2} \sigma_S^2 \theta^2 v_{zz} + \theta \sigma_S \sigma_F \rho v_{zf} = 0.
\end{equation}
This is the \emph{dynamic programming equation} for the value function $v$. As the optimal portfolio \eqref{eq:feedback} depends on $v$ and its derivatives, this partial differential equation is fully nonlinear. Nevertheless, it can be solved in closed form for standard utility functions of power or exponential form and a number of particular asset dynamics \cite{merton.71,kim.omberg.96,chacko.viceira.05,liu.07}. This in turn leads to explicit expressions for the optimal trading policy \eqref{eq:feedback}. This is illustrated in Section~\ref{sec:examples} for the model with mean-reverting returns first studied by \cite{kim.omberg.96}; the closed-form solution for the corresponding dynamic programming equation~\eqref{eqn:frictionlessoperator} is provided in Appendix~\ref{sec:KOValueFunction}.

\subsection{Dynamic Programming with Proportional Transaction Costs}\label{ss:prop}

\begin{figure}
	\centering
    \begin{tikzpicture}
	\matrix{
	    \node at (0,0) {\includegraphics[width=0.48\textwidth]{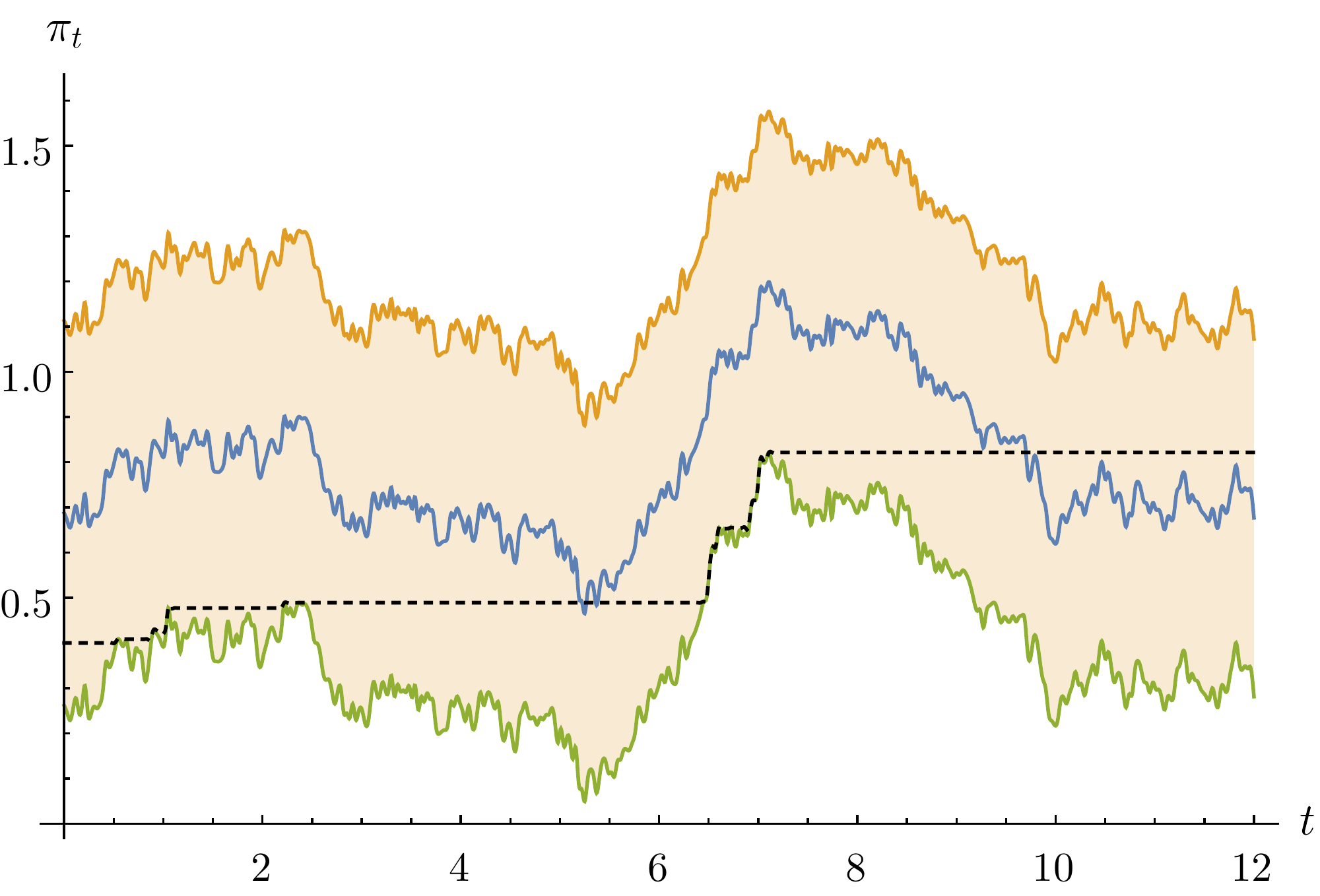}};
	    &
	    \node at (0,0) {\includegraphics[width=0.48\textwidth]{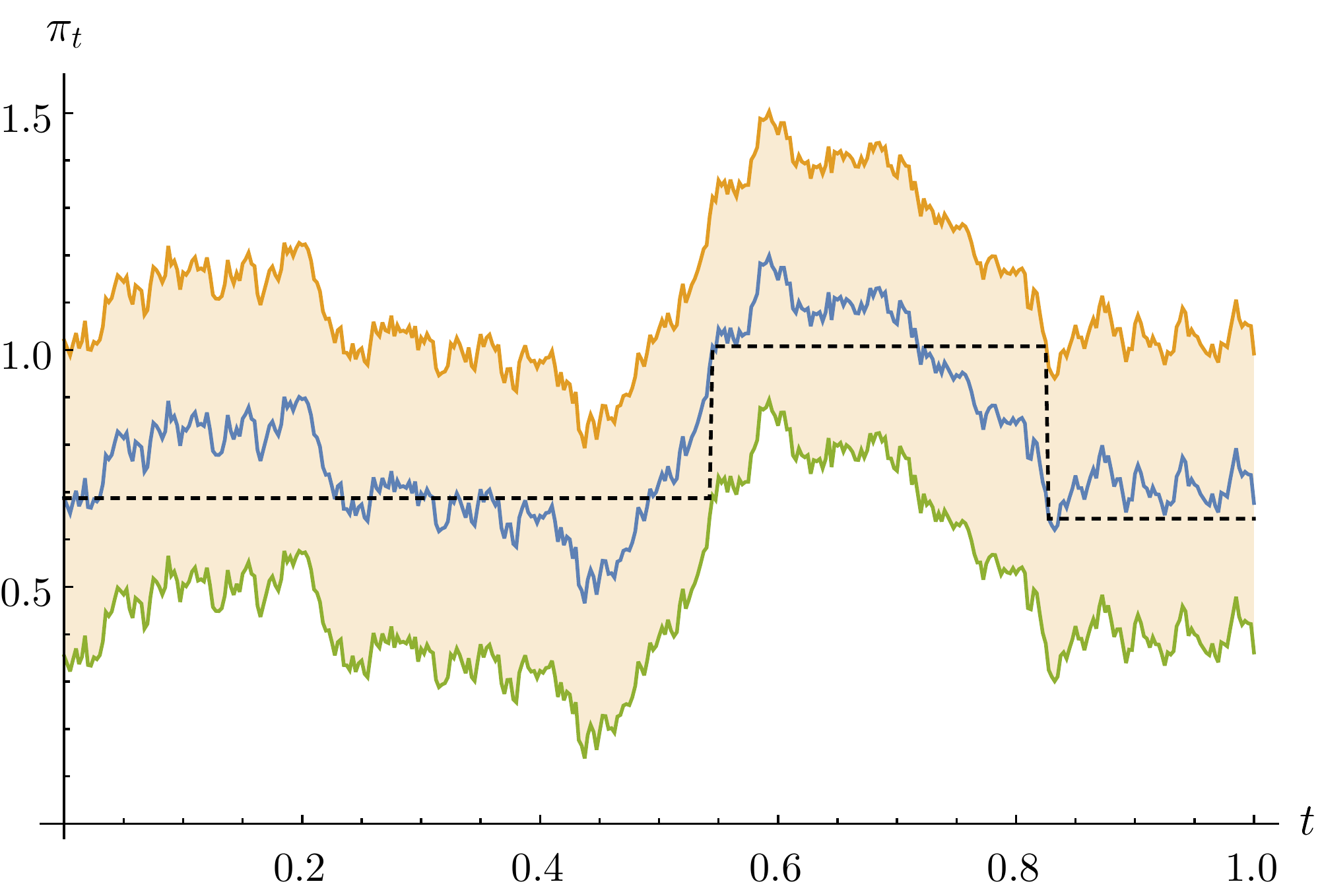}}; \\
	};
    \end{tikzpicture}
	
    \caption{\label{fig:NTsimulation} Simulation of the frictionless portfolio weight $\pi_t=Y_t/(X_t+Y_t)$ and the boundaries of the corresponding (asymptotic) no-trade regions in the Kim--Omberg model, in solid lines. The dashed lines depict the paths of optimal frictional portfolios for proportional transaction costs of $1\%$ (left panel) and for fixed costs of \$$1$ for an investor with an initial wealth of \$$5000$ (right panel). The parameters are $\gamma = 3$, $r=0.0168$, $\sigma_S = 0.151$, $\kappa = 0.271$, $\bar{F} = 0.041$ and $\sigma_F = 0.0343$.}
\end{figure}

Let us now pass to the value function \eqref{eq:vf2} with proportional transaction costs. In this case, 
the value function depends not only on the total wealth $z$ but rather
on the full portfolio decomposition $(x,y)$, where $x$ is 
the cash position and $y$ is the dollar amount invested in the risky asset. The dynamic programming principle then takes the following form:
 \begin{align*}
     v^\lambda(t,x,y,f)& = \,\, \sup_{\mathclap{\dif{M}_t,\dif{L}_t}} \,\, E_t\left[v^\lambda\left(t+\dif{t}, x+\dif{X}^{t,x,y,f}_{t},y+\dif{Y}^{t,x,y,f}_{t},f+\dif{F}^{t,f}_t\right)\right].
 \end{align*}
Like in the frictionless case, apply It\^o's formula, insert the state dynamics (\ref{eq:facdyn}, \ref{e.cash}--\ref{e.risky}), cancel the stochastic integrals (because they are martingales and therefore have zero expectation if the integrands are sufficiently integrable), divide by $\dif{t}$, and send $\dif{t}$ to zero. This leads to 
\begin{multline}
    0 = v^\lambda_t + \mu_F v^\lambda_f   +
y(r+\mu_S) v^\lambda_y +x r v^\lambda_x +\frac{ \sigma_F^2}{2} v^\lambda_{ff}
+ \frac{\sigma_S^2 y^2}{2} v^\lambda_{yy}  +\rho \sigma_S\sigma_F y v^\lambda_{yf} \\
\quad  + \,\, \sup_{\mathclap{\dif{M}_t,\dif{L}_t}} \,\, \Bigg\{ \left(v^\lambda_y-(1+\lambda) v^\lambda_x\right)\frac{\dif{L}_t}{\dif{t}}+\left((1-\lambda) v^\lambda_x - v^\lambda_y\right)\frac{\dif{M}_t}{\dif{t}}\Bigg\}.
\label{eq:dpe2}
\end{multline}
Hence, if the marginal utility of increasing the risky position is neither too high nor too low at a point $(t,x,y,f)$,
\[  (1+\lambda) v^\lambda_x(t,x,y,f)  > v^\lambda_y(t,x,y,f)> (1-\lambda) v^\lambda_x(t,x,y,f), \]
then it is not optimal to transact at all ($\dif{M}_t=\dif{L}_t=0$). 
We call the set of all such positions $(t,x,y,f)$ the \emph{no-trade region}.
In view of the dynamic programming equation \eqref{eq:dpe2}, it follows that the following standard linear 
PDE is satisfied inside it:
\begin{equation}\label{eq:linearPDE}
\begin{split}
    0 &= v_t + \mu_F v_f   + y(r+\mu_S) v_y +x r v^\lambda_x +\frac{ \sigma_F^2}{2} v_{ff}
    + \frac{\sigma_S^2 y^2}{2} v_{yy}  +\rho \sigma_S\sigma_F y v_{yf}.
\end{split}
\end{equation}
Outside the no-trade region, \eqref{eq:dpe2} shows that the right-hand side of this equation is less than or equal to zero, since one can always let the portfolio evolve uncontrolled. Moreover, if the marginal utility of increasing the risky position is high enough, $v_y \geq (1+\lambda)v_x$, then the optimal control is to ``buy risky shares at an infinite rate'' ($\dif{L_t}/\dif{t}=\infty$). Conversely, it is optimal to ``sell risky shares at an infinite rate'' ($\dif{M_t}/\dif{t}=\infty$) whenever this marginal utility is low enough, $v_y \leq (1-\lambda)v_x$. This means that it is optimal to perform the minimal amount of trading that keeps the portfolio within the no-trade region, cf.\ the left panel of Figure~\ref{fig:NTsimulation} for an illustration. Moreover, it follows that in order to satisfy the dynamic programming equation \eqref{eq:dpe2} with equality, we must have $v_y = (1+\lambda)v_x$ in the ``buying region'' and $v_y = (1-\lambda)v_x$ in the ``selling region''. Together with \eqref{eq:linearPDE}, this leads to the following ``variational inequality'' for the value function \eqref{eq:vf2}:
\begin{equation}\label{eq:viprop}
\begin{aligned}
    0 = \max \Big\{ v^\lambda_t + \mu_F v^\lambda_f   + y(r+\mu_S) v^\lambda_y +x r v^\lambda_x +\frac{\sigma_F^2}{2} v^\lambda_{ff}
	+ \frac{\sigma_S^2 y^2}{2} v^\lambda_{yy}  +\rho \sigma_S\sigma_F y v^\lambda_{yf}\ ;& \\
(1+\lambda)v^\lambda_x - v^\lambda_y \ ;\  v^\lambda_y -(1-\lambda)v^\lambda_x &\Big \} .
\end{aligned}
\end{equation}
Explicit solutions for this equation are not available even in the simplest models. The key difficulty is that the transaction costs increase the number of state variables by one and introduce a free boundary---the no-trade region is unknown and needs to be determined as part of the solution. In the Black--Scholes model, this requires to solve for a time-dependent smooth curve \cite{dai.yi.09}.\footnote{In the most tractable infinite horizon models, the trading boundaries are constant and can be characterized by a scalar nonlinear equation \cite{dumas.luciano.91,taksar.al.88,gerhold.al.13,gerhold.al.14}; see \cite{guasoni.muhlekarbe.13} for a survey of this literature.} In the Kim--Omberg model, the tractability issue is further compounded, because the trading boundaries then additionally depend on the mean-reverting expected  return process. This lack of analytical tractability can be overcome by passing to the small-cost limit, which we discuss in Section~\ref{sec:homogenization}.

\subsection{Dynamic Programming with Fixed Costs}
\label{ss.dpfixed}

Before turning to the small-cost asymptotics, let us briefly sketch how to adapt the above derivations for models with additional fixed costs. Since only finitely many trades $\Delta L_t>0$ or $\Delta M_t>0$ are possible, we have
\[  v^\lambda(t,x,y,f) \ge v^\lambda(t, x- \Delta L_t+\Delta M_t-\lambda_p (\Delta L_t+\Delta M_t) - \lambda_f, y+ \Delta L_t-\Delta M_t, f).  \]
With
$$
H(t,x,y, v^\lambda(t, \cdot,\cdot,f)):= \,\sup_{\mathclap{\ell, m \ge 0}} \,
\left\{v^\lambda\left(t, x- \ell+m-\lambda_p (\ell+m) - \lambda_f, 
y+ \ell-m, f\right)\right\},
$$
it follows that the frictional value function $v^\lambda$ satisfies the following inequality:
$$v^\lambda(t,x,y,f)\ge H(t,x,y, v^\lambda(t, \cdot,\cdot,f)).$$
The states for which $v^\lambda >H$ again form a no-trade region, where the same argument as in Section~\ref{ss:prop} shows that the frictional value function solves the linear PDE \eqref{eq:linearPDE}. Combining this with the nonlinear constraint $v^\lambda=H$ which is binding outside the no-trade region, the following variational inequality for the frictional value function is obtained:
\begin{align*}
    0 = \max \Big\{ v^\lambda_t + \mu_F v^\lambda_f + y(r+\mu_S) v^\lambda_y +x r v^\lambda_x+ \frac{\sigma_F^2}{2} v^\lambda_{ff}
	+ \frac{\sigma_S^2 y^2}{2} v^\lambda_{yy}  +\rho \sigma_S\sigma_F y v^\lambda_{yf} \ ;& \\
    H(t,x,y, v^\lambda(t, \cdot,\cdot,f)) -v^\lambda(t,x,y,f) &\Big \} .
\end{align*}
With only fixed costs ($\lambda_p=0$), all trades are penalized equally, so that the corresponding optimal policy rebalances all the way back to the frictionless target, as shown in the right panel of Figure~\ref{fig:NTsimulation}. Models with both fixed and proportional costs are intermediate between this regime and that of proportional costs from Section~\ref{ss:prop} in the sense that the optimal policy is to trade to a point in-between the boundary of the no-trade region and the frictionless target portfolio, compare \cite{korn.98,altarovici.al.16}.

%% file: homogenization.tex
\section{Homogenization}
\label{sec:homogenization}

We now turn to the asymptotic analysis of models with \emph{small} transaction costs. To ease the exposition, we focus a single risky asset traded with purely proportional costs. A multi-asset model with proportional costs is discussed in \cite{possamai.al.15}; fixed and proportional costs are treated in \cite{altarovici.al.15,altarovici.al.16}; a study of quadratic trading costs can be found in \cite{moreau.al.15}. 

As already pointed out above, explicit solutions for portfolio choice problems with transaction costs are not available even in those settings that can be solved in closed form in the frictionless case. To overcome this lack of tractability, it is natural to study small transaction costs as a perturbation of the frictionless benchmark. The goal is to ``reveal the salient features of the problem while remaining a good approximation to the full but more complicated model'' \cite{whalley.wilmott.97}.

The method developed in \cite{soner.touzi.13} that we present here has its roots in the 
homogenization literature \cite{papanicolaou27boundary,kozlov1979averaging}. 
This class of problems contains an ergodic fast variable and the theory studies
the limit problem as this variable oscillates more and more quickly.  This leads to a ``homogenized equation''.  Interestingly, the latter is not simply
the ergodic average of the original one.  Instead, it is obtained by a non-trivial 
coupling with a so-called \emph{corrector equation} (sometimes also called the \emph{cell equation}).
Models with small transaction costs are only loosely in analogy with
these problems as the dependence on the portfolio 
composition disappears in the limit and it
does not immediately offer a fast variable.
However, \cite{soner.touzi.13} observed that---after a suitable rescaling---the deviation of
the portfolio from the target position, ($\xi$ in \eqref{eqn:fastvariable} below)
plays the same role as the fast variable.  This observation allows
to employ the similar formal calculations as in homogenization theory to characterize the asymptotic solution.
Moreover, the powerful perturbed test function method of Evans \cite{evans1992periodic,evans1989perturbed}
can be modified to obtain rigorous convergence results \cite{soner.touzi.13,possamai.al.15,moreau.al.15,altarovici.al.15,altarovici.al.16,bouchard2014hedging}.

\subsection{Identifying the Correct Scalings}\label{ss:scalings}

The starting point for the asymptotic analysis is an appropriate ansatz for the value function $v^\lambda$ with small transaction costs $\lambda$. To this end, a key observation of \cite{janecek.shreve.04,rogers.04} is that two competing effects needs to be balanced here. On the one hand, a narrower no-trade region leads to more frequent trading, and whence also higher direct transaction costs. On the other hand, a wider no-trade region leads to larger oscillations around the frictionless target portfolio and thus higher indirect losses due to displacement from the optimal risk-return trade-off.

With proportional transaction costs, the amount of trading required to remain inside a small no-trade region with width $\Delta$ scales with the inverse of $\Delta$.\footnote{This is a property of reflected Brownian motion and the local time it accumulates at the boundaries.} Locally around the frictionless optimum, the first-order condition implies that value function is quadratic, so that the displacement loss should scale with the squared width $\Delta^2$ of the no-trade region. In summary, this suggests that $\Delta$ needs to minimize a total loss of the form
\[  C_1 \Delta^2 + \frac{C_2 \lambda}{\Delta}. \]
As a consequence, the optimal no-trade region should be of order $\mathcal{O}(\lambda^{1/3})$ with a corresponding minimal utility loss of order $\mathcal{O}(\lambda^{2/3})$.\footnote{For fixed costs, the argument is similar: A trade is initiated whenever the portfolio reaches the boundary of the no-trade region. At such a point, the portfolio is rebalanced to the frictionless portfolio. The time it takes Brownian motion to reach the boundary of the no-trade region is proportional to $\Delta^2$, so the number of trades per unit of time is proportional to $1/\Delta^2$. The corresponding utility loss should therefore be of the form $C_1 \Delta^2 + C_2 \lambda/\Delta^2$, which has a minimizer of order $\mathcal{O}(\lambda^{1/4})$ and a minimal value of order $\mathcal{O}(\lambda^{1/2})$; cf.~\cite{altarovici.al.15} for more details.}

\subsection{Ansatz for the Asymptotic Frictional Value Function}

As discussed in Section~\ref{ss:prop}, the frictional value function $v^\lambda$ depends on time $t$, the current value $f$ of the factor process, as well as the current safe and risky positions $x$ and $y$. As the transaction cost $\lambda$ tends to zero, both the risky and safe position converge to their frictionless counterparts. In order to obtain nontrivial limiting quantities for the asymptotic analysis, we therefore re-parametrize the model by switching from $x$ and $y$ to the frictionless state variable
\[  z=x+y\]
and the \emph{normalized} deviation
\begin{equation} \label{eqn:fastvariable}
    \xi = \frac{y - \theta(t,z,f)}{\lambda^{1/3}}
\end{equation}
of the risky position from its frictionless target \eqref{eq:feedback}.%
{\footnote{The definition of this fast variable depends on the 
scaling appropriate for the problem at hand.
For example, for problems with fixed rather than proportional costs, one needs
to divide by $\lambda^{1/4}$, cf.~\cite{altarovici.al.15}.}} In view of the discussion in Section~\ref{ss:scalings}, we then expect \eqref{eqn:fastvariable} to converge to a finite limit as $\lambda \to 0$. To avoid fractional powers in the calculations below, set
\[  \epsilon = \lambda^{1/3}.\]
With this notation and the above change of variables, the frictional value function can be written as 
\[  v^\lambda(t,x,y,f) =: v^\epsilon(t,z,\xi,f).\]
The considerations in Section~\ref{ss:scalings} suggest that the leading-order term in the asymptotic expansion of the value function is of order $\mathcal{O}(\epsilon^2)$. Since the impact of a single trade is of higher order $\mathcal{O}(\epsilon^3)$, this term should not depend on the deviation \eqref{eqn:fastvariable} and should thus be a function $\epsilon^2 u(t,z,f)$ of the frictionless state variables only. However, the deviation of the frictionless optimizer (i.e., the position in the no-trade region) evidently plays a key role in determining the optimal trading policy (i.e., when to start trading). In order to take this into account and motivated by the homogenization literature, we introduce a second term $\epsilon^4 w(t,z,\xi,f)$ in the asymptotic expansion. It is negligible at the leading order in the value expansion, but via \eqref{eqn:fastvariable} its derivatives play a crucial role in determining the optimal trading policy from the variational inequality \eqref{eq:viprop}.

In summary, our ansatz for the asymptotic value function reads as follows:
\begin{equation} \label{eqn:expansion}
  v^\epsilon(t,z,\xi,f)= v(t,z,f) - \epsilon^2 u(t,z,f) - \epsilon^4 w(t,z,\xi,f) + \mathcal{O}(\epsilon^3).
\end{equation}
The goal now is to determine $u$ and $w$ by plugging this expansion into the dynamic programming equation \eqref{eq:viprop} and matching terms sorted in powers of the asymptotic parameter $\lambda = \epsilon^3$.

\subsection{Asymptotic Dynamic Programming and Corrector Equations}

In order to recast the variational inequality \eqref{eq:viprop} in terms of the new variables $z$ and $\xi$ instead of $x$ and $y$, we need to rewrite the corresponding differential operators. For an arbitrary function $\Psi$ of $(t,x,y,f)$, or equivalently $(t,z,\xi,f)$, we have
\begin{align*}  
D_x \Psi(t,x,y,f) &= D_z \Psi(t,z,\xi,f) - \frac{\theta_z(t,z,f)}{\epsilon} D_\xi \Psi(t,z,\xi,f), \\
D_y \Psi(t,x,y,f) &= D_z \Psi(t,z,\xi,f) + \frac{1 - \theta_z(t,z,f)}{\epsilon} D_\xi \Psi(t,z,\xi,f),
\end{align*}
and in turn
\begin{align*} 
D_{yy} \Psi(t,x,y,f) &=D_z \left(D_z \Psi(t,z,\xi,f) + \frac{1 - \theta_z(t,z,f)}{\epsilon} D_\xi \Psi(t,z,\xi,f)\right) \\
&\quad + \frac{1 - \theta_z(t, z, f)}{\epsilon} D_\xi \left(D_z \Psi(t,z,\xi,f) + \frac{1 - \theta_z(t,z,f)}{\epsilon} D_\xi \Psi(t,z,\xi,f)\right). 
\end{align*}
Likewise,
\begin{align*}
    D_f \Psi(t,x,y,f) &= D_f \Psi(t,z,\xi,f) - \frac{\theta_f}{\epsilon} D_\xi \Psi(t,z,\xi,f), \\
    D_{ff} \Psi(t,x,y,f) &= D_f \left(D_f \Psi(t,z,\xi,f) - \frac{\theta_f(t,z,f)}{\epsilon} D_\xi \Psi(t,z,\xi,f)\right) \\
	&\quad - \frac{\theta_f(t,z,f)}{\epsilon} D_\xi \left( D_f \Psi(t,z,\xi,f) - \frac{\theta_f(t,z,f)}{\epsilon} D_\xi \Psi(t,z,\xi,f) \right), \\
    D_{yf} \Psi(t,x,y,f) &= D_z \left(D_f \Psi(t,z,\xi,f) - \frac{\theta_f(t,z,f)}{\epsilon} D_\xi \Psi(t,z,\xi,f)\right) \\
	&\quad + \frac{1 - \theta_z(t,z,f)}{\epsilon} D_\xi\left(D_f \Psi(t,z,\xi,f) - \frac{\theta_f(t,z,f)}{\epsilon} D_\xi \Psi(t,z,\xi,f) \right).
\end{align*}
Note the terms of order $\mathcal{O}(\epsilon^{-2})$ arising in some of these expressions. These are the reason why $\epsilon^4 w(t,z,\xi,f)$ cannot be absorbed in $\mathcal{O}(\epsilon^3)$ in \eqref{eqn:expansion} but needs to be treated separately.

With the above expressions, the ansatz \eqref{eqn:expansion} implies
\begin{equation} \label{eqn:expansionderivatives}
    \begin{aligned}
	D_x v^\epsilon &= D_z v - \epsilon^2 D_z u + \epsilon^3 \theta_z w_\xi + \mathcal{O}(\epsilon^4) = D_z v - \epsilon^2 D_z u + \mathcal{O}(\epsilon^3), \\
	D_y v^\epsilon &= D_z v - \epsilon^2 D_z u - \epsilon^3 (1- \theta_z) w_\xi + \mathcal{O}(\epsilon^4) = D_z v - \epsilon^2 D_z u + \mathcal{O}(\epsilon^3), \\
	D_{yy} v^\epsilon &= D_{zz} v - \epsilon^2 \left(D_zz u + (1 - \theta_z)^2 D_{\xi\xi} w \right) + \mathcal{O}(\epsilon^3), \\
	D_f v^\epsilon &= D_f v - \epsilon^2 D_f u + \mathcal{O}(\epsilon^3), \\
	D_{ff} v^\epsilon &= D_{ff} v - \epsilon^2 \left( D_{ff}u + \theta_f^2 D_{\xi\xi}w \right) + \mathcal{O}(\epsilon^3), \\
	D_{yf} v^\epsilon &= D_{zf} v - \epsilon^2\left( u - (1 - \theta_z) \theta_f D_{\xi\xi}w \right) + \mathcal{O}(\epsilon^3).
    \end{aligned}
\end{equation}
We want to use these expressions to expand the frictional dynamic programming equation \eqref{eq:viprop}. To this end, recall its frictionless counterpart \eqref{eqn:frictionlessoperator},
\[  \mathcal{A} v := v_t + \mu_F v_f + \frac{1}{2} \sigma_F^2 v_{ff} + r z v_z + \mu_S \theta v_z + \frac{1}{2} \sigma_S^2 \theta^2 v_{zz} + \theta \sigma_S \sigma_F \rho v_{zf} = 0, \]
where the corresponding optimal risky position is
\begin{equation} \label{eqn:thetadef}
    \theta = \frac{\mu_S v_z + \rho \sigma_S \sigma_F v_{fz}}{-\sigma_S^2 v_{zz}}.
\end{equation}
Together with \eqref{eqn:expansionderivatives}, and using $y = \theta + \epsilon \xi = \theta + \mathcal{O}(\epsilon)$ to replace $\epsilon^2 y$ with $\epsilon^2 \theta + \mathcal{O}(\epsilon^3)$, the PDE \eqref{eq:linearPDE} in the no-trade region can now be expanded as follows:
\begin{equation} \label{eqn:NTexpanded}
    \begin{aligned}
	\mathcal{L} v^\epsilon &:= v^\epsilon_t + \mu_F v^\epsilon_f + y(v^\epsilon_y (\mu_S + r) - v^\epsilon_x r) + rz v^\epsilon_x + \frac{1}{2} \sigma_S^2 y^2 v^\epsilon_{yy} + \sigma_S y \sigma_F \rho v^\epsilon_{fy} + \frac{1}{2} \sigma_F^2 v^\epsilon_{ff} \\
	&\phantom{:}= \underbrace{\mathcal{A}v}_{\stackrel{\eqref{eqn:frictionlessoperator}}{=}0} + \underbrace{(y - \theta) \mu_S v_z + \frac{1}{2} \sigma_S^2 (y^2 - \theta^2) v_{zz} + \sigma_S (y - \theta) \sigma_F \rho v_{zf}}_{I} \\
	&\quad - \epsilon^2\underbrace{\left(u_t + r z u_z+ \mu_S y u_z + \mu_F u_f + \frac{1}{2} \sigma_F^2 u_{ff} + \frac{1}{2} \sigma_S^2 y^2 u_{zz} + \sigma_S y \sigma_F \rho u_{zf} \right) }_{II} \\
	&\quad - \epsilon^2 w_{\xi\xi} \frac{1}{2} \underbrace{\left( \sigma_S^2 \theta^2 (1 - \theta_z)^2 - 2 \sigma_S \sigma_F \rho \theta (1-\theta_z) \theta_f + \sigma_F^2 \theta_f^2 \right)}_{\mathclap{\alpha^2}} \\
	&\quad + \mathcal{O}(\epsilon^3)
    \end{aligned}
\end{equation}
By \eqref{eqn:thetadef},
\begin{align*}
    I = \left(\mu_S v_z + \sigma_S \sigma_F \rho v_{fz}\right)(y - \theta) + \frac{1}{2} \sigma_S^2 (y^2 - \theta^2)&= - \sigma_S^2 v_{zz} \theta (y - \theta) + \frac{1}{2} \sigma_S^2 (y^2 - \theta^2) \\
    &= \frac{1}{2} \sigma_S^2 v_{zz} (y - \theta)^2 \\
    &= \frac{1}{2} \sigma_S^2 v_{zz} \epsilon^2 \xi^2.
\end{align*}
Next, note that $y = \theta + \epsilon \xi = \theta + \mathcal{O}(\epsilon)$ implies%
\footnote{Recall that $\mathcal{A}$ is a \emph{nonlinear} operator in the frictionless dynamic programming equation \eqref{eqn:frictionlessoperator}, because the frictionless control $\theta$ depends on the solution $v$ of the equation. In contrast, $\theta$ is already determined in terms of $v$ here, so that $\mathcal{A}$ acts as a \emph{linear} operator on $u$.}
\[  II = \mathcal{A} u + \mathcal{O}(\epsilon). \]
In summary, \eqref{eqn:NTexpanded} simplifies to the following asymptotic expansion of the dynamic programming equation in the no-trade region:
\[  \mathcal{L} v^\epsilon = - \epsilon^2 \left( - \frac{1}{2} \sigma_S^2 \xi^2 v_{zz} + \mathcal{A}u + \frac{1}{2} \alpha^2 w_{\xi\xi}\right)+ \mathcal{O}(\epsilon^3). \]

It remains to derive expansions in the buy and sell regions. To this end, we rewrite the gradient constraint from \eqref{eq:viprop}, \[  v^\epsilon_x - (1 - \epsilon^3)v^\epsilon_y, \]
using the expressions from \eqref{eqn:expansionderivatives}, obtaining 
\[  v^\epsilon_x - (1-\epsilon^3)v^\epsilon_y = \epsilon^3 v_y + \underbrace{(v_x - v_y)}_{\epsilon^3 w_\xi + \mathcal{O}(\epsilon^4)} = \epsilon^3 (v_z + w_\xi) + \mathcal{O}(\epsilon^4). \]
Analogously, the second gradient constraint in \eqref{eq:viprop} can be expanded as follows:
\[ v^\epsilon_y - (1 - \epsilon^3)v^\epsilon_x = \epsilon^3(v_z - w_\xi) + \mathcal{O}(\epsilon^4). \]
Altogether, the asymptotic dynamic programming equation is
\[ \min \left\{ \epsilon^2 \left( - \frac{1}{2} \sigma_S^2 \xi^2 v_{zz} + \mathcal{A}u + \frac{1}{2} \alpha^2 w_{\xi\xi}\right); \epsilon^3 (v_z + w_\xi ); \epsilon^3(v_z - w_\xi) \right\} = 0. \]
Since factoring out $\epsilon^2$ and $\epsilon^3$ does not change this equation, it  is equivalent to
\begin{equation}\label{eq:viasy}
\min \left\{ - \frac{1}{2} \sigma_S^2 \xi^2 v_{zz} + \mathcal{A}u + \frac{1}{2} \alpha^2 w_{\xi\xi}; v_z + w_\xi; v_z - w_\xi \right\} = 0.
\end{equation}

The variational inequality \eqref{eq:viasy} with two unknowns $w$ and $u$ turns out to effectively consist of two separate equations. To see why, write
\[  a(t,z, f) := \mathcal{A}u(t,z, f). \]
Then \eqref{eq:viasy} can be rewritten as an equation for $w$ and $a$:
\begin{equation}\label{eq:wa}
 \min \left\{ - \frac{1}{2} \sigma_S^2 \xi^2 v_{zz} + a + \frac{1}{2} \alpha^2 w_{\xi\xi}; v_z + w_\xi; v_z - w_\xi \right\} = 0,
\end{equation}
where
\begin{equation} \label{eqn:alphasq}
	\alpha^2 := \sigma_S^2 \theta^2 (1 - \theta_z)^2 - 2 \sigma_S \sigma_F \rho \theta (1-\theta_z) \theta_f + \sigma_F^2 \theta_f^2,
\end{equation}
is determined by the model parameters and the frictionless optimizer. For each value of $t$, $z$ and $f$, \eqref{eq:wa} has a solution $\xi \mapsto w(t,z,\xi,f)$ for precisely one value of $a=a(t,z,f)$. Thus, by finding the solution $(w,a)$ to this equation, we have obtained a unique function $a(t,z,f)$, from which we can in turn determine $u$ as the solution of a \emph{linear} PDE:

\[  \mathcal{A}u = a. \]
The key to this separation is the uniqueness of the solution $(w, a)$ to \eqref{eq:wa}.%
\footnote{For any solution $(w,a)$, $(w+C,a)$ is also a solution for any constant $C$, so the uniqueness only concerns $a$. However, for a given choice of normalization, e.g., $w(z,0) = 0$, also $w$ is uniquely determined. The choice of normalization affects neither the equation for $u$, nor the policy generated from $w$. A precise formulation of the uniqueness result is presented in~\cite[Theorem 3.1]{possamai.al.15}. Similar uniqueness results are proven in \cite{hynd.12,menaldi.al.13}.}
Any other choice of function $u'$ would give another value $a' = \mathcal{A}u'$, for which, by uniqueness of $(w,a)$, there would not exist a solution $w'$. 

In summary, the asymptotic expansion \eqref{eqn:expansion} of the frictional value function is determined by the following equations:

\begin{definition}[First corrector equation]
    For any $(t,z,f)$, the \emph{first corrector equation} for the pair $(w(t,z,\cdot,f), a(t,z,f))$ is 
    \begin{equation}\label{eqn:firstcorrector}
	\min \Bigg\{ \underbrace{- \frac{1}{2} \sigma_S^2 \xi^2 v_{zz} + a + \frac{1}{2} \alpha^2 w_{\xi\xi}}_{\text{no trade region}}; \underbrace{\vphantom{\frac{1}{2}}v_z + w_\xi; v_z - w_\xi}_{\text{trade regions}} \Bigg\} = 0.
    \end{equation}
    Since any constant can be added to a solution $w$ to obtain another solution, we will impose the normalization $w(t,z,0,f) = 0$ which affects neither the value expansion nor the optimal policy at the leading asymptotic order.
\end{definition}

\begin{definition}[Second corrector equation]
    Given a solution $(w, a)$ of the first corrector equation, the \emph{second corrector equation} for $u(\cdot,\cdot,\cdot)$ is
    \[  \mathcal{A}u(t,z,f) = a(t,z,f), \quad\quad u(T,z,f) = 0, \]
    where $\mathcal{A}$ is the generator of the frictionless optimal wealth process, appearing in \eqref{eqn:frictionlessoperator}. \end{definition}
    
 The intuition for this separation into two equations is the following. In the first corrector equation, only the deviation $\xi$ of the portfolio from its frictionless target is a variable. In contrast, the frictionless state variables $(t,z,f)$ are treated as constants because they vary much more slowly than $\xi$ for small transaction costs. Conversely, the ``fast  variable'' $\xi$ is averaged out in the second corrector equation determining the leading order utility loss $u$, in that it does not enter directly but only through the function $a(t,z,f)$ computed from the first corrector equation.

\subsection{Explicit Solutions}\label{ss:expl}

In the present setting, the first corrector equation can be solved explicitly. This allows us to understand the comparative statics of the asymptotically optimal no-trade region and the corresponding leading-order welfare effect of small transaction costs. As a byproduct, the calculations below also explain why there is only a single value of $a$ for which the first corrector equation has a solution. 

To find the solution, fix $(t,z,f)$ and make the ansatz that (i) the no-trade region is a symmetric interval $(-\Delta\xi, \Delta\xi)$ around the frictionless optimizer ($\xi=0$), and (ii) the asymptotic value function is of the following form:\footnote{This is the lowest order symmetric polynomial in the deviation $\xi$ with enough degrees of freedom to ensure value matching and smooth pasting at the trading boundaries $\pm\Delta\xi$.}
\[
    w(\xi) = w(t,z,\xi,f) = 
    \begin{cases}
	c_4 \xi^4 + c_2 \xi^2 &\text{if } |\xi| \leq \Delta\xi, \\
	w(\Delta\xi) + (\xi - \Delta\xi) & \text{if } \xi \geq \Delta\xi, \\
	w(\Delta\xi) - (\xi - \Delta\xi) & \text{if } \xi \leq \Delta\xi.
    \end{cases}
\]
Here, $c_4$, $c_2$ are parameters to be determined along with $a$ and $\Delta\xi$. In the no-trade region, plugging this ansatz into \eqref{eqn:firstcorrector} leads to
$$
  a = \frac{1}{2} \sigma_S^2 \xi^2 v_{zz} - \alpha^2 (6 c_4 \xi^2 + c_2) = \left(\frac{1}{2} \sigma_S^2 v_{zz} - 6\alpha^2 c_4 \right) - \alpha^2 c_2. 
$$
Since this needs to hold for any value $\xi \in (-\Delta \xi,\Delta\xi)$, comparison of the coefficients of $\xi^2$ and $1$ yields
\begin{equation}\label{eq:ci}
 c_4 = \frac{\sigma_S^2 v_{zz}}{12 \alpha^2} \quad \text{and} \quad c_2 = -\frac{a}{\alpha^2}.
 \end{equation}
To pin down $a$ and $\Delta\xi$, we impose that the value function is not only continuous but also twice continuously differentiable across the trading boundaries $\pm\Delta\xi$.\footnote{These are the ``smooth pasting conditions'' of \cite{benes.al.80,dumas.91}.} By symmetry, this leads to the following two additional conditions:
\[
    \begin{aligned}
	0 &= 12 c_4 (\Delta\xi)^2 + 2 c_2, \\
	v_z &= 4 c_4 (\Delta\xi)^3 + 2 c_2 \Delta\xi.
    \end{aligned}
\]
These equations readily yield
\begin{equation}\label{eq:a}
a = \frac{\sigma_S^2 v_{zz}}{2}\Delta \xi^2,
\end{equation}
with
\begin{equation} \label{eqn:xi0}
    \Delta\xi = \Delta\xi(t,z,f) = \left( \frac{-v_z}{v_{zz}} \frac{3 \alpha^2}{2 \sigma_S^2}  \right)^{\frac{1}{3}}.
 \end{equation}   
 Together with \eqref{eq:ci}, this leads to a closed-form solution of the first corrector equation \eqref{eqn:firstcorrector} in terms of model parameters and inputs from the frictionless optimization problem.
    
\subsection{Asymptotically Optimal Policy}    
    
Recalling that $\lambda = \epsilon^3$, we find that the asymptotically optimal no-trade region corresponding to the leading-order variational inequality \eqref{eq:viasy} is 
\begin{equation}\label{eq:ntapprox}
  \text{NT}_\lambda \approx \{ (t,x,y,f) : \abs{y - \theta(t,x+y,f)} \leq \lambda^{1/3} \Delta\xi(t,z,f) \}.
\end{equation}
In view of \eqref{eqn:xi0} and the representation \eqref{eqn:alphasq} for $\alpha^2$, this asymptotic no-trade region is determined by (i) the diffusion coefficients $\sigma_S$, $\sigma_F$ of the risky asset and the factor process, (ii) the frictionless optimal portfolio $\theta$ and its derivatives $\theta_z$, $\theta_f$, and (iii) the risk-tolerance $-v_z/v_{zz}$ of the frictionless value function. The comparative statics of this formula for general utilities are discussed in \cite{kallsen.muhlekarbe.15}. Here, we focus on the case most relevant for applications: power utilities with constant relative risk aversion.

\paragraph{Constant Relative Risk Aversion}
As is well known (cf., e.g., \cite{zariphopoulou.01}), power utilities $U(x)=x^{1-\gamma}/(1-\gamma)$ imply that the optimal \emph{fraction} of wealth invested in the risky asset, $\pi(t,f):=\theta(t,z,f)/z$, is independent of the wealth level. Moreover, the value function inherits the homotheticity $v(t,z,f)=z^{1-\gamma}v(t,1,f)$, so $-\frac{v_{zz}}{v_{z}}(t,z,f)=\gamma/z$. In view of \eqref{eqn:xi0} and \eqref{eqn:alphasq}, the asymptotic no-trade region can therefore be written in terms of risky weights (in contrast to monetary amounts) as
\begin{equation}\label{eq:ntapproxpower} 
\text{NT}_\lambda \approx \left\{ (t,x,y,f) : \abs{\frac{y}{x+y} - \pi(t,f)} \leq \lambda^{1/3} \Delta\pi(t,f) \right\}, 
\end{equation}
where
\begin{equation}\label{eq:frac}
    \Delta \pi=\left(\frac{3}{2\gamma}\left(\pi^2(1-\pi)^2-2\pi(1-\pi)\pi_f \frac{\sigma_F}{\sigma_S}+\pi^2_f \frac{\sigma^2_F}{\sigma^2_S}\right)\right)^{\mathrlap{1/3}}.
\end{equation}
Hence, in this case, the halfwidth of the asymptotically optimal no-trade region is fully determined by the volatilities $\sigma_S$, $\sigma_F$, the risk aversion $\gamma$, as well as the frictionless portfolio weight $\pi$ and sensitivity $\pi_f$ with respect to the state variable $f$. 

\subsection{Welfare loss}\label{sec:esr}

With the explicit solution of the first corrector equation from Section~\ref{ss:expl}, we can also say more about the leading order term $\epsilon^2u(t,z,f)$ in the expansion \eqref{eqn:expansion} of the frictional value function. To this end, recall that $\mathcal{A}u=a$, where $a$ is given by \eqref{eq:a} and the differential operator $\mathcal{A}$ is defined in \eqref{eqn:frictionlessoperator}. Now, note that $\mathcal{A}$ is the infinitesimal generator of the frictionless optimal wealth process $Z_t$. Whence, the Feynman--Kac formula and \eqref{eq:a} show that
\begin{equation}\label{eq:cel1}
\begin{split}
    u(t,z,f)&=E_t\left[ \int_t^T -a(s,Z_s,F_s)\dif{s}\right]\\
   &= v_z(t,z,f) E^Q_t\left[\int_t^T  -\frac{v_{zz}(s,Z_s,F_s)}{v_z(s,Z_s,F_s)}\frac{\sigma_S(F_s)^2}{2}\Delta \xi(s,Z_s,F_s)^2 \dif{s}\right], 
\end{split}
\end{equation}
where $Q$ is the frictionless investor's ``marginal pricing measure'', whose density process is proportional to the marginal indirect utility $v_z$ by the first-order condition of convex duality \cite{davis.97,karatzas.kou.96}.\footnote{In complete markets, $Q$ is simply the unique equivalent martingale measure. In any case, the corresponding density process is known explicitly if the value function of the problem at hand can be computed in closed form.}

With the representation \eqref{eq:cel1} for $u$, Taylor's theorem allows to rewrite the expansion \eqref{eqn:expansion} as
\begin{equation}
    \label{eqn:celexpansion}
    v^\lambda (t, z, f) = v\left(t, z - \mathrm{CEL}^\lambda(t, x, f), f\right)+\mathcal{O}(\lambda).
\end{equation}
Here,
\begin{equation}
    \label{eqn:cel}
    \mathrm{CEL}^\lambda (t, z, f) = \lambda^{2/3} E^Q_t\left[\int_t^T  -\frac{v_{zz}(s,Z_s,F_s)}{v_z(s,Z_s,F_s)}\frac{\sigma_S(F_s)^2}{2}\Delta \xi(s,Z_s,F_s)^2 \dif{s}\right]
\end{equation}
is the \emph{certainty equivalent loss} due to small transaction costs---the amount of initial endowment the investor would forego in order to trade the risky asset without transaction costs. Like the asymptotic no-trade region, this measure for the welfare effect of the trading costs is determined by (i) the diffusion coefficients of the risky asset and the factor process, (ii) the frictionless optimal policy and its sensitivities, and (iii) the risk tolerance of the frictionless value function. As all of these quantities are generally random and time dependent, they are averaged both with respect to time and states.

\paragraph{Constant Relative Risk Aversion}

For power utilities with constant relative risk aversion $\gamma>0$, Formula~\eqref{eqn:cel} can be further simplified. To wit, we then have
\begin{equation}
    \label{eqn:cel2}
    \mathrm{CEL}^\lambda (t, z, f) = z \lambda^{2/3}E^{\tilde{P}}_t \left[ \int_t^T \frac{\gamma}{2} \sigma_S(F_s)^2 \Delta \pi(s,F_s)^2 \dif{s} \right],
\end{equation}
where the expectation is computed under the measure $\tilde{P}$ whose density process is proportional to the value function $v(s,Z_s,F_s)$ evaluated along the frictionless optimal wealth process.\footnote{This measure also plays an important role in the asymptotic analysis of small unhedgeable risks, compare \cite{kramkov.sirbu.06}.} By scaling out the current wealth $z$, this leads to the \emph{relative certainty equivalent loss}---an appealing scale-invariant measure for the welfare effect of transaction costs, also used in the numerical work of \cite{balduzzi.lynch.99}, for example. Here, its small-cost approximation is obtained by averaging the frictionless optimizer, its sensitivities, and the volatilities of the risky asset and factor process in a suitable way. These are all functions of the factor process $F$. To find its dynamics under the measure $\tilde{P}$, use It\^o's formula to compute the dynamics of the density process:
\begin{equation}\label{eq:density}
    \frac{\dif{v(s, Z_s, F_s)}}{v(s, Z_s, F_s)} = \frac{\mathcal{A}v(s, Z_s, F_s)}{v(s, Z_s, F_s)} \dif{s} + \frac{v_z(s, Z_s, F_s)}{v(s, Z_s, F_s)} \dif{Z_s} + \frac{v_f(s, Z_s, F_s)}{v(s, Z_s, F_s)} \dif{F_s}=:\dif{L}_t,
\end{equation}
The $\dif{s}$-term vanishes by the frictionless dynamic programming equation~\eqref{eqn:frictionlessoperator}. It follows that the density process of $\tilde{P}$ is a stochastic exponential $\mathcal{E}(L)_t=\exp(L_t-\frac{1}{2}\langle L\rangle_t)$, and the $\tilde{P}$ dynamics of $F$ can be readily computed using Girsanov's theorem. If the frictionless value function is known in closed form, the corresponding change of drift is once more fully explicit.

\paragraph{Long Investment Horizons}

The relative certainty equivalent loss \eqref{eqn:cel2} is a function of time and the state variable only. If the planning horizon $T$ is long, the time variable averages out and even more tractable formulas obtain. To wit, for large $T$, the frictionless policy $\pi(t,f)$ typically quickly converges to a steady-state value $\bar{\pi}(f)$ that only depends on the state variable (but not the current time), and the frictionless value function approximately scales as follows \cite{guasoni.robertson.12}:
\begin{equation}\label{eqn:esr}
    v(t,z,f) \approx \frac{z^{1-\gamma}}{1-\gamma}  e^{(1-\gamma)(T-t)\mathrm{ESR}}.
\end{equation}
Here, the \emph{equivalent safe rate} $\mathrm{ESR}$ is a fictitious interest rate which---in the long run---yields the same growth rate of utility as trading in the original market. With small transaction costs, \eqref{eqn:celexpansion}, \eqref{eqn:cel2}, and \eqref{eqn:esr} show that the corresponding leading-order expression is\footnote{This requires that the effect of transaction costs is small even when compounded over a long horizon. To make this argument precise, one can directly consider the infinite-horizon problem as in \cite{gerhold.al.14,kallsen.muhlekarbe.15,melnyk.seifried.15}.}
\[  v^\lambda(t,x,y) \approx  \frac{z^{1-\gamma}}{1-\gamma} e^{(1-\gamma)(T-t)\left(\mathrm{ESR}-\Delta \mathrm{ESR}_t^\lambda\right)}. \]
Here, the \emph{equivalent safe rate loss} due to small transaction costs is
\[  \Delta\mathrm{ESR}_t^\lambda= \lambda^{2/3} \frac{\gamma}{2}\frac{1}{(T-t)} E_t^{\tilde{P}}\left[\int_t^T (\Delta\pi(s_,F_s)^2  \sigma_S(F_s)^2 \dif{s} \right]. \]
On any finite time horizon $T$, this quantity is a function of time $t$ and the value $f$ of the state variable. However, as the horizon grows, the ergodic theorem suggests that if the state variable $F$ has a stationary distribution $\nu_F^{\tilde{P}}(\dif{f})$ under the measure $\tilde{P}$, then the equivalent safe rate loss converges to a constant, like the frictionless equivalent safe rate:\footnote{We denote by $\Delta\bar{\pi}(f)$ the halfwidth of the stationary no-trade region obtained from the long-run portfolio $\bar{\pi}(f)$ via \eqref{eq:frac}.}
\begin{equation}\label{eqn:esrsteady}
    \Delta\mathrm{ESR}^\lambda \approx \lambda^{2/3} \frac{\gamma}{2}  \int_{-\infty}^\infty \Delta\bar\pi(f)^2  \sigma(f)^2 \nu_F^{\tilde{P}}(\dif{f}).
\end{equation}
In summary, the welfare effect in infinite-horizon models with small transaction costs can be computed by performing a simple numerical quadrature.

%% file: examples.tex
\section{Examples}
\label{sec:examples}
\newcommand{\Fbar}{\bar{F}}

We now illustrate the asymptotic results from Section~\ref{sec:homogenization} in two concrete examples. As a sanity check, we first consider the Black--Scholes model and verify that the formulas derived above indeed coincide with the expressions directly obtained for this simple model (cf., e.g., \cite{janecek.shreve.04,bichuch.12,gerhold.al.14,soner.touzi.13}). Afterwards, we turn to the model of Kim and Omberg, and discuss how the results change with stochastic investment opportunities.

\subsection{Black--Scholes Model}

For an investor with constant relative risk aversion $\gamma>0$, the value function in the Black--Scholes model is $v(t,z)=\frac{z^{1-\gamma}}{1-\gamma} \exp\left((1-\gamma)(r+\frac{\mu^2}{2\gamma\sigma^2})(T-t)\right)$
and the corresponding risky weight is constant: $\pi_{BS}=\mu/\gamma\sigma^2$. As a consequence, the halfwidth \eqref{eq:frac} of the asymptotic no-trade region simplifies to
$$\Delta\pi_{BS}= \left(\frac{3}{2\gamma}\pi_{BS}^2(1-\pi_{BS})^2\right)^{\mathrlap{1/3}}.$$
The corresponding formula for the leading-order equivalent safe rate loss is 
\[ \Delta\mathrm{ESR}^\lambda = \frac{\gamma\sigma^2}{2} \left(\frac{3\lambda}{2\gamma}\pi_{BS}^2(1-\pi_{BS})^2\right)^{\mathrlap{2/3}}. \]
Both of these expressions vanish if zero or full investment is optimal in the frictionless model. Then, the respective frictionless optimal strategies never trade, and no transaction costs need to be paid. In contrast, transaction costs play a more important role if the frictionless target weight is close to $1/2$, but even then the quantitative effects are rather small \cite{constantinides.86}. If a leveraged portfolio is optimal ($\pi_{BS}>1$), the welfare loss can be more substantial \cite{gerhold.al.14}.

\subsection{Kim--Omberg Model}
Let us now sketch how the above results change in the Kim--Omberg model, where the expected excess return follows an Ornstein--Uhlenbeck process with dynamics \eqref{eq:oudyn}. For a power utility function with constant relative risk aversion $\gamma>1$, the frictionless value function $v$ has the following closed-form expression~\cite{kim.omberg.96}:
\begin{equation}
    \label{eqn:KOv}
    v(s,z,f) = \frac{z^{1-\gamma}}{1-\gamma} \exp \left(A(s) + B(s)f + \frac{1}{2} C(s) f^2 \right),
\end{equation}
where the functions $A$, $B$, and $C$ are the explicit solutions of some Riccati equations~\cite{kim.omberg.96}.\footnote{For the convenience of the reader, we recall these formulas in Appendix~\ref{sec:KOValueFunction}.} The corresponding optimal risky weight is linear in the state variable:
\[	\pi_{KO}(s,F_s) = \frac{F_s}{\gamma \sigma_S^2} + \frac{\rho \sigma_F}{\gamma \sigma_S} ( B(s) + C(s) F_s ). \]
In view of \eqref{eq:frac}, this formula immediately yields a closed-form expression for the halfwidth $\Delta \pi_{KO}$ of the asymptotic no-trade region with transaction costs. Figure \ref{fig:NTsimulation} shows a simulated sample path of the frictionless portfolio $\pi_{KO}$ and the boundaries of the no-trade region $\pi_{KO} \pm \Delta\pi_{KO}$. 

\begin{figure}
	\centering
	\includegraphics[width=0.7\textwidth]{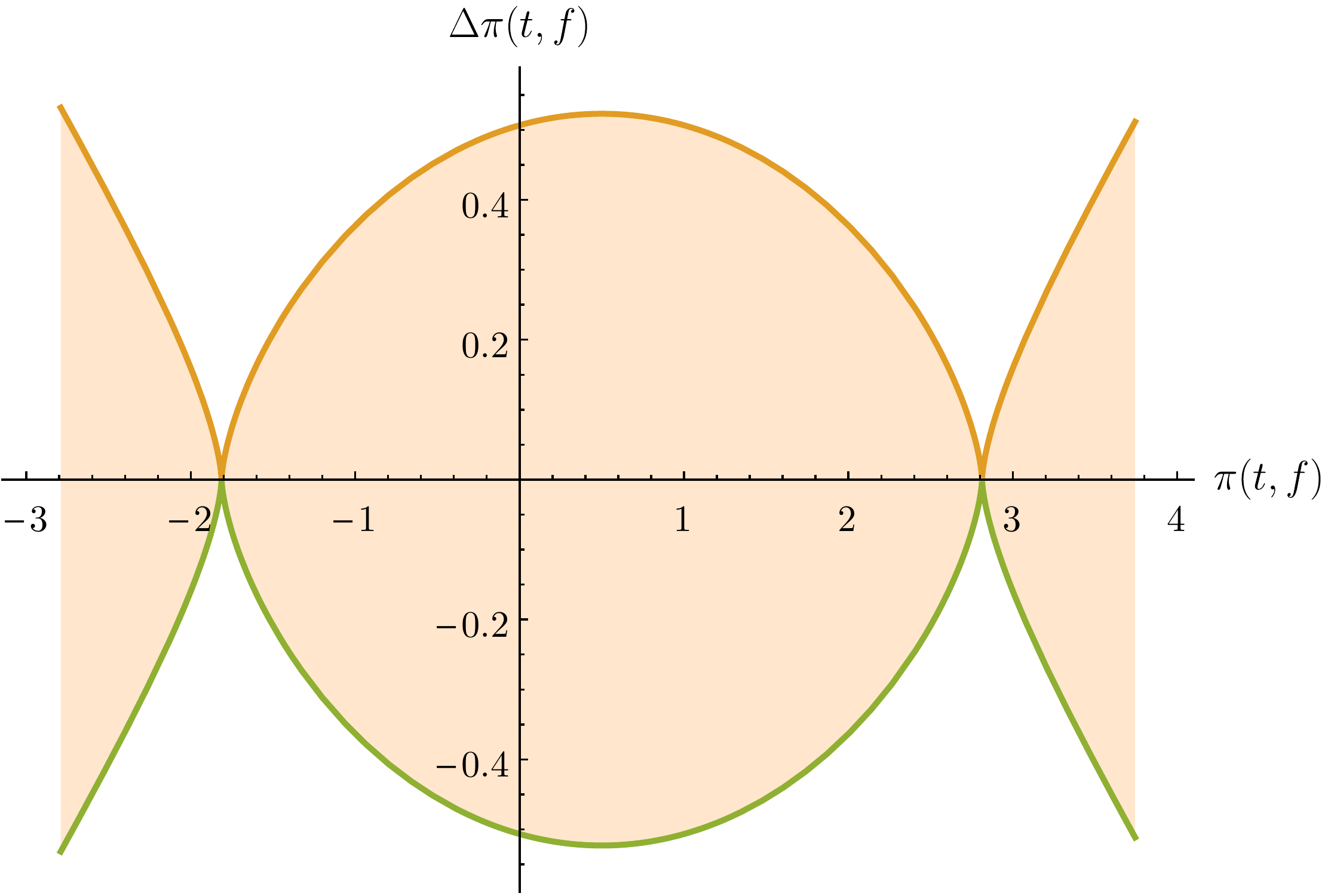}
	\caption{\label{fig:delta-pi-VS-pi} Halfwidth of the no-trade region in the Kim--Omberg model plotted against the optimal frictionless risky weight. (Yearly) Parameters are $T=40$, $\gamma = 3$, and $r=0.0168$, $\sigma_S = 0.151$, $\kappa = 0.271$, $\Fbar = 0.041$, $\sigma_F = 0.0343$ as in \cite{barberis.00}. }
\end{figure}

Since the frictionless risky weight is sensitive to changes in the state variable, the no-trade region no longer vanishes if the frictionless risky weight is zero or one, but instead at two other levels determined by the model parameters. For the parameters estimated from a long equity time series in \cite{barberis.00}, this is illustrated in Figure \ref{fig:delta-pi-VS-pi}. There, the halfwidth of the no-trade region is plotted against the optimal frictionless risky weight for different values of the expected excess return.

\begin{figure}
	\centering
	\includegraphics[width=0.7\textwidth]{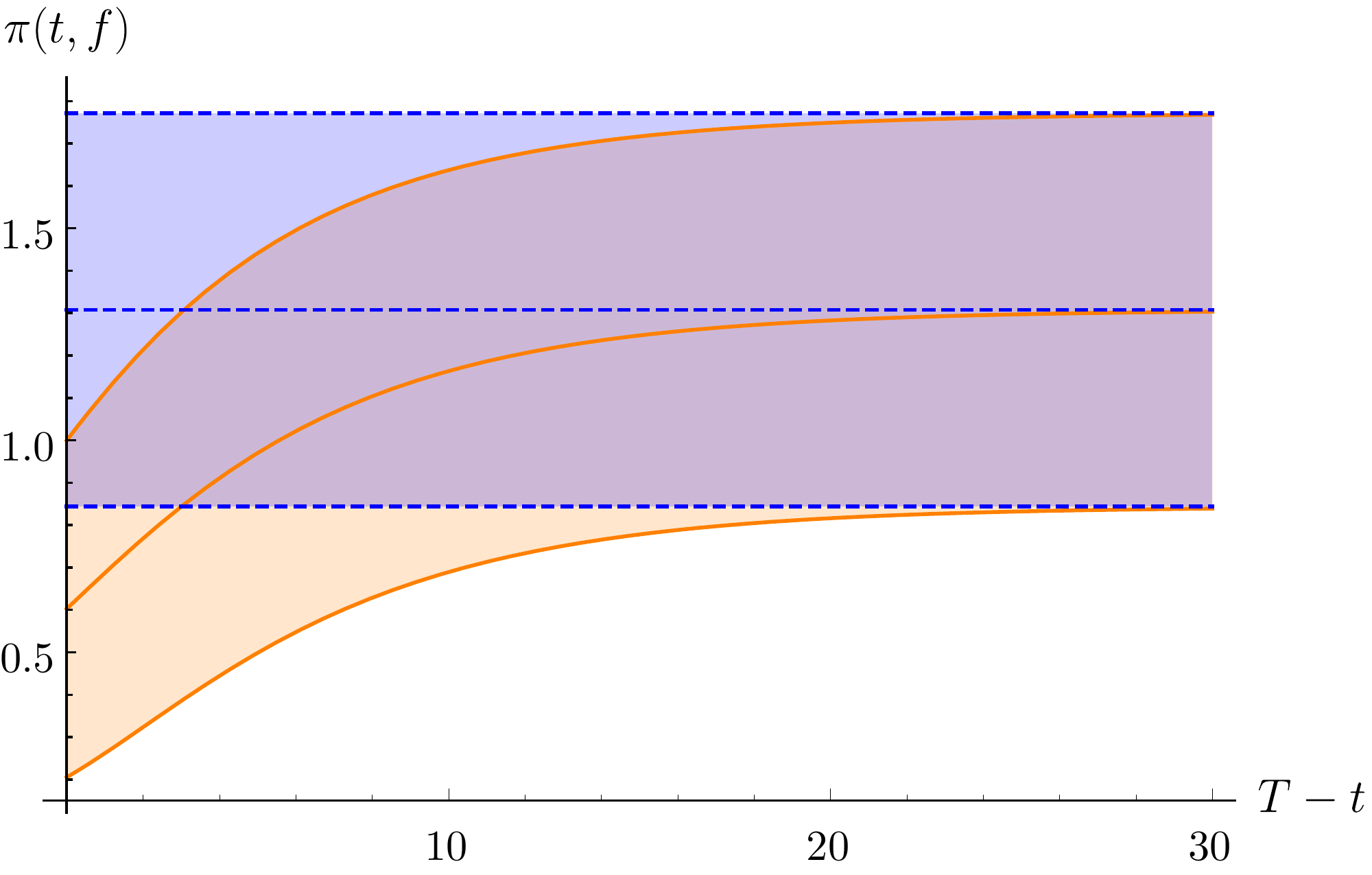}
	\caption{\label{fig:longHorizonConvergence} The frictionless position and no-trade region (solid) as a function of time (measured in years), plotted alongside the infinite-horizon values (dashed). (Yearly) Parameters are $\gamma = 3$ and $r=0.0168$, $\sigma_S = 0.151$, $\kappa = 0.271$, $\Fbar = 0.041$, $\sigma_F = 0.0343$ as in \cite{barberis.00}. }
\end{figure}

This is complemented by Figure~\ref{fig:longHorizonConvergence}, which shows how the optimal frictionless portfolio and the corresponding no-trade region converge to their stationary long-run limits as the planning horizon grows. This stationary policy $\bar{\pi}_{KO}$ corresponds to the stationary points $\bar{B}$ and $\bar{C}$ of the Riccati equations for $B(s)$ and $C(s)$. The corresponding no-trade region is in turn derived from \eqref{eq:frac}.

\begin{figure}
	\centering
	\includegraphics[width=0.7\textwidth]{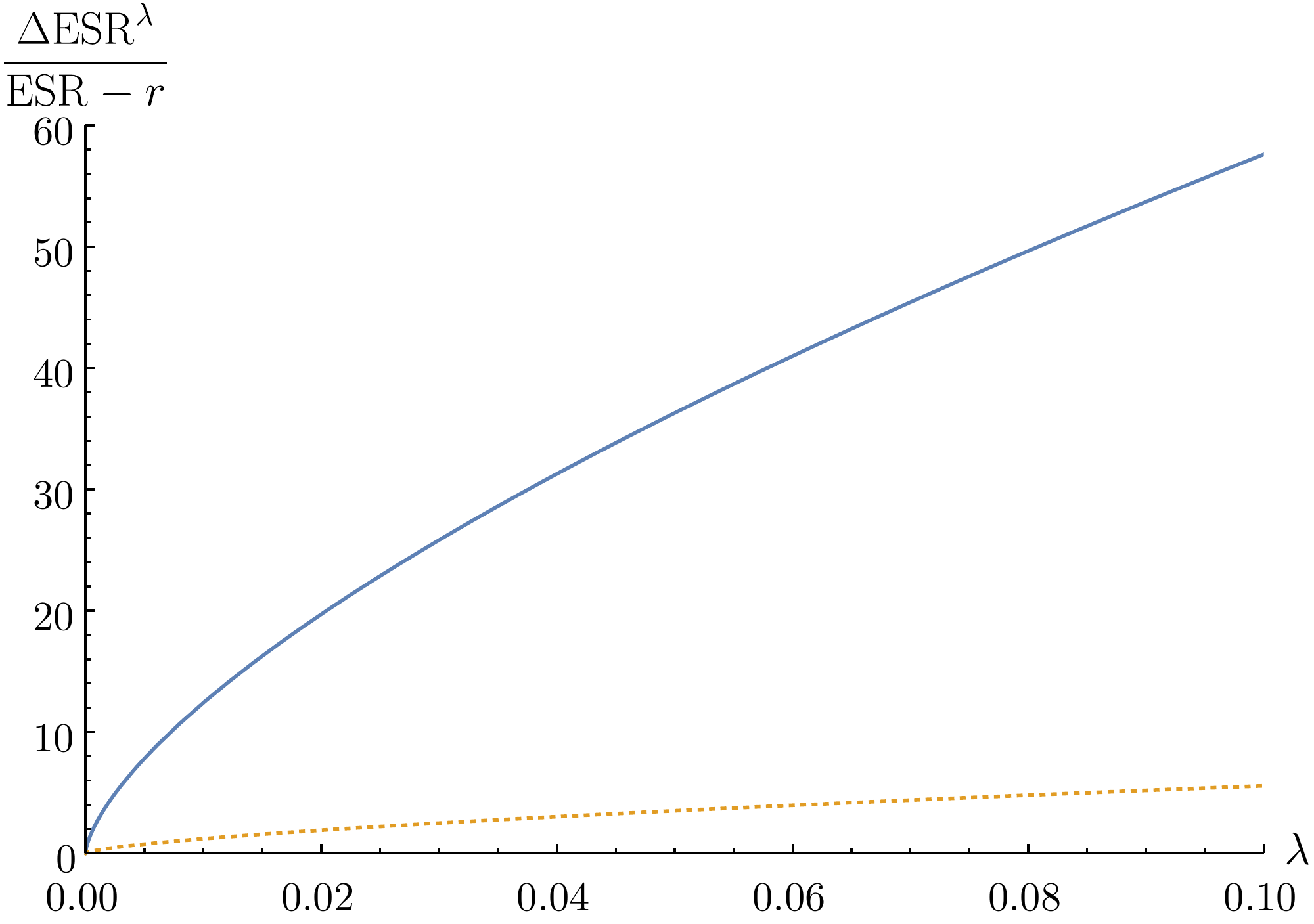}
	\caption{\label{fig:ESR-epsilon}Relative loss in equivalent safe rate due to transaction costs for the Kim--Omberg model (solid) and a Black--Scholes model (dotted) with $\mu_S \equiv \Fbar$, plotted against the size of the cost. (Yearly) Parameters are $\gamma = 3$ and $r=0.0168$, $\sigma_S = 0.151$, $\kappa = 0.271$, $\Fbar = 0.041$, $\sigma_F = 0.0343$ as in \cite{barberis.00}.}
\end{figure}

The long-horizon convergence of the functions $B(s)$ and $C(s)$ can also be used to simplify the computation of the leading-order relative certainty equivalent loss \eqref{eqn:esrsteady}. In particular, Girsanov's theorem and the long-run convergence show that the long term dynamics of the factor process $F$ under the measure $\tilde{P}$ with density process \eqref{eq:density} are
\begin{align*}
    \dif{F}_s &= \kappa (\Fbar - F_s) \dif{s} + \sigma_F \dif{W}^F \\
    &= \big( \kappa(\Fbar - F_s) + \sigma_F (1 - \gamma) \pi_{KO}(s,F_s) \sigma_S \rho + (B(s) + C(s) F_s) \sigma_F^2 \big) \dif{s} + \sigma_F \dif{\tilde{W}^F} \\
    &\approx \big( \kappa(\Fbar - F_s) + \sigma_F (1 - \gamma) \bar{\pi}_{KO}(F_s) \sigma_S \rho + (\bar{B} + \bar{C} F_s) \sigma_F^2 \big) \dif{s} + \sigma_F \dif{\tilde{W}^F} \\
    &=: \tilde{\kappa} ( \tilde{F} - F_s) \dif{s} + \sigma_F \dif{\tilde{W}^F},
\end{align*}
for a $\tilde{P}$-Brownian motion $\tilde{W}^F$ and suitably chosen constants $\tilde{\kappa}$ and $\tilde{F}$. Hence, for a long planning horizon, the factor process still has Ornstein--Uhlenbeck dynamics under the auxiliary measure $\tilde{P}$, and its stationary law is $\nu_F^{\tilde{P}} \sim \mathcal{N}(\tilde{F}, \sigma_F^2 / 2 \tilde{\kappa})$. This allows computation of the relative certainty equivalent loss \eqref{eqn:esrsteady} due to small transaction costs according to
\begin{equation}
    \label{eqn:esrsteadyKO}
    \Delta\mathrm{ESR}_{KO}^\lambda \approx \lambda^{2/3} \frac{\gamma}{\sqrt {2 (\sigma_F^2/\tilde{\kappa})^{2}\pi }}  \int_{-\infty}^\infty \left(\Delta\bar\pi_{KO}(f)^2  \sigma(f)^2 \exp\left({-{\frac {(f-\tilde{F} )^{2}}{2(\sigma_F^2/2\tilde{\kappa})^{2}}}}\right)\right) \dif{f},
\end{equation}
Since the integrand is known in closed form, \eqref{eqn:esrsteadyKO} is easily evaluated by numerical quadrature. This is illustrated in Figure~\ref{fig:ESR-epsilon} which plots the relative certainty equivalence loss as a function of the proportional transaction cost. Compared to a Black--Scholes model with the same expected  excess return, we observe that the welfare effect of transaction costs is indeed increased substantially by having to react to the time-varying investment opportunities. 

%% file: extensions.tex
\section{Extensions}
\label{s.extension}

So far, we have focused on the application of the homogenization approach to a portfolio choice problem with proportional transaction costs for purchases and sales of a single risky asset with Markovian dynamics. Performance was measured in terms of expected utility from terminal wealth only, i.e., intertemporal consumption was absent. This choice was made for concreteness and ease of exposition. In this section, we survey results from the recent literature that show that the findings outlined in Section~\ref{sec:homogenization} remain true much more generally. 

\subsection{More General Preferences}
\label{ssec:generalprefs}

The results of the previous sections readily extend to models with intermediate consumption. For example, \cite{soner.touzi.13} is a study of an infinite-horizon model with preferences of the form
$$E\left[\int_0^\infty e^{-\delta t} U(c_t) \dif{t}\right] \to \max!,$$
where $\delta>0$ is the investor's discount rate and $U(c_t)$ measures the utility from consumption (rate) $c_t$ at time $t$. The asymptotic no-trade region for this optimization criterion turns out to be of the same form as in \eqref{eq:ntapprox}---intermediate consumption is only reflected through the frictionless optimal policy. This remains true for more general ``additive'' preferences of the form
\begin{equation}\label{eq:additive}
    E\left[\int_0^T U_1(t,c_t) \dif{t}+U_2(Z_T)\right] \to \max!, 
\end{equation}
where $U_1(t,c_t)$ is the utility from intermediate consumption at time $t \in [0,T]$ and $U_2(Z_T)$ is the utility from terminal wealth $Z_T$ at time $T$; see \cite{kallsen.muhlekarbe.15,ahrens.15} for more details. Despite this robustness result, substantial intertemporal consumption can have a nonnegligible effect if the trading costs are not small enough. The intuition is that since costs are paid from the savings account, investors are willing to accept smaller risky positions before rebalancing \cite{davis.norman.90}. In infinite-horizon Black--Scholes models, this manifests itself through a downward shift of the no-trade region at the \emph{second} asymptotic order $O(\lambda^{2/3})$, see~\cite{janecek.shreve.04,lo.al.04,gerhold.al.12}. For more general models, such results are not available. 

The effects of the transaction costs on the optimal consumption policy are of the simplest conceivable form: it is asymptotically optimal to simply adjust the frictionless rule for the (typically lower) wealth with transaction costs \cite{kallsen.muhlekarbe.15}. For power utility, this implies that the consumption/wealth ratio is unaffected by the trading costs \cite[Section 4]{kallsen.muhlekarbe.15}.

Very recently, it has been shown \cite{melnyk.al.16} that even the additive structure in the preferences \eqref{eq:additive} is not crucial, in that the same leading-order results also remain true for recursive preferences as in \cite{duffie.epstein.92} or models with habit formation such as \cite{hindy.huang.93}. At the leading asymptotic order, the optimal trading strategy is again completely characterized by the local curvature of the agents' preferences, measured by the risk-tolerance of their indirect utilities, and the consumption wealth ratio remains unchanged. The fine structure of the preferences at hand only enters at the next-to-leading order.

\subsection{More General Asset Dynamics}

In previous sections, we have assumed that the joint dynamics (\ref{eq:risky}--\ref{eq:facdyn}) of the asset prices and a factor process are Markovian, i.e., all drift and diffusion coefficients are deterministic functions of the current state of the system. This assumption allows to apply PDE techniques, but is not crucial for the validity of formulas \eqref{eq:ntapprox} and \eqref{eqn:celexpansion}. Indeed, these formulas are derived by ``freezing'' the frictionless state variables for the analysis of the first corrector equation---a procedure that readily generalizes to general, not necessarily Markovian systems where drift and diffusion coefficients can be arbitrary functionals of the information flow. Even in such more general models, formulas \eqref{eq:ntapprox} for the asymptotically optimal no-trade region and \eqref{eqn:celexpansion} for the leading-order effect of small transaction costs remain valid, see \cite{kallsen.muhlekarbe.15a,kallsen.muhlekarbe.15,cai.al.15,cai.al.16} for more details. 

Extending the approach presented here to models with jumps leads to the analysis of nonlocal integro-differential equations. While this appears to be a daunting task, it is shown in \cite{rosenbaum.tankov.14}, using probabilistic techniques, that asymptotic results similar to \eqref{eq:ntapprox} and \eqref{eqn:celexpansion} can still be obtained.

\subsection{More General Transaction Costs}

Although we have focused on proportional transaction costs here, also this structure is not crucial. At least on a formal level, fixed costs \cite{altarovici.al.15}, fixed and proportional costs \cite{altarovici.al.16}, or quadratic costs \cite{moreau.al.15} can be treated similarly.

The fine structure of the optimal strategies crucially depends on the cost at hand. With proportional costs, one performs the minimal amount of trading to remain in a no-trade interval around the frictionless target. With fixed costs, it is no longer possible to implement such a strategy involving infinitely many small trades; hence, one directly trades back to a target portfolio once the boundaries of the no-trade region is reached. Conversely, quadratic costs lead to smaller penalties for very small trades but make large turnover rates prohibitively expensive. Thus, optimal strategies always trade towards the target at some finite, absolutely continuous rate. 

Despite these apparent differences, the ``coarse'' structure of all these models is nevertheless very similar: In each case, the distance from the target is a trade-off against the specific trading cost, balanced by an appropriate control. The corresponding expected displacement and average transaction costs display the same comparative statics in each case, up to a change of asymptotic convergence rates and constants. In particular, the implications of transaction costs for welfare and average trading volume are very similar in each case. See \cite{moreau.al.15} for more details.

\subsection{Multiple Risky Assets}

The extensions sketched so far eventually lead to explicit asymptotic formulas of a similar complexity as for the benchmark model discussed in Section \ref{sec:homogenization}.  In contrast, less is known about the case of several risky assets. In this case, the homogenization approach still reduces the dimensionality of the problem \cite{possamai.al.15} but the resulting corrector equations no longer admit an explicit solution. As a consequence, numerical methods such as the policy iteration scheme in Section \ref{sec:policy_iteration} are needed even for the asymptotic analysis. Models with quadratic costs \cite{garleanu.pedersen.13,garleanu.pedersen.16,guasoni.weber.13,moreau.al.15} and fixed costs \cite{atkinson.wilmott.95,altarovici.al.15} can still be solved in closed form in the multidimensional case, but the tractability issue is only exacerbated for general nonlinear costs.

%% file: numerics.tex
\section{Numerical solutions in multiple dimensions}
\label{sec:policy_iteration}

The corrector equations obtained by passing to the small-cost limit are considerably simpler than the original dynamic programming equation. Even in situations with multiple risky assets \cite{possamai.al.15} or nonlinear costs \cite{altarovici.al.16} where explicit solutions are not available, this simplifies the numerical analysis considerably. In this section, we present a variant of the classical ``policy iteration algorithm'' which is tailored to the singular control problem at hand and works very well in practice.\footnote{Alternatively methods, based on PDE techniques, can be found in \cite{muthuraman.kumar.06, dai.zhong.10}.}
An attractive feature of this method is that both the value function and the corresponding no-trade region are approximated simultaneously. 

The main idea of policy iteration is to start with a guess for the optimal control, and compute the corresponding payoff by solving a linear equation. The result of this computation is then used to generate an ``improved'' policy  by---in each point---determining the best response to this payoff. These two steps are in turn iterated until a fixed point is found, see, e.g., \cite{puterman} for an overview.

To apply this technique in the presence of proportional transaction costs, the first step is to recognize that the first corrector equation can be interpreted as the dynamic programing equation of an ergodic control problem. Instead of using a ``direct'' policy iteration schemes like in \cite{chancelier.al.07}, we approximate the singular controls by smooth trading rates, capped at some finite level (compare \cite[Section 3]{davis.norman.90} and \cite{witte.reisinger.12}), to achieve better stability (cf., e.g., \cite{azimzadeh.forsyth.16}). The resulting control problem can then be discretized and solved by a classical scheme. The approximation by capping is a special case of the approach presented in \cite{altarovici.al.16} for problems with proportional and fixed costs, and can readily generalized to other settings.

Consider a $d$-dimensional model with frictionless value function $v$ and optimal risky positions $\theta$. To simplify notation,  all assets are traded with the same proportional transaction cost. Analogous arguments to those in Section~\ref{sec:homogenization} then show that the $d$-dimensional version of the first corrector equation~\eqref{eqn:firstcorrector} is
\begin{equation}\label{eq:ergodic}
    \min_{i = 1,\dots,d} \min \Bigg\{ - \frac{1}{2} |\sigma_S^\top \xi|^2 v_{zz} + a + \frac{1}{2} \Tr \left[ \alpha\alpha^\top w_{\xi\xi} \right];  \;\;
	v_z + w_\xi \cdot e_i; \;\;
	v_z - w_\xi \cdot e_i
    \Bigg\} = 0,
\end{equation}
where $\alpha$ is a model-dependent matrix analogous to \eqref{eqn:alphasq}, depending on $\theta$ and its derivatives.

The corrector equation \eqref{eq:ergodic} can be interpreted as the dynamic programming equation of an infinite-horizon control problem. To wit, let $L^i$, $M^i$, $i=1,\dots,d$ be nondecreasing controls for 
\[  \Xi^i_s = \xi^i + \sum_{j = 1}^d \alpha^{i,j}(t,z,f) W^j_s + L^i_s - M^i_s. \]
Then, \eqref{eq:ergodic} is the dynamic programming equation for the ergodic control problem 
\begin{equation}\label{eq:ergodic2}
a(t,z,f) := \inf_{L,M} J(t,z,f;L,M),
\end{equation}
corresponding to the following infinite-horizon goal functional:
\[  J(t,z,f; L, M) := \limsup_{s \rightarrow \infty} \frac{1}{s} \E \left[ \frac{1}{2} \int_0^s -v_{zz}(t,z,f) \left| \sigma_S^\top \Xi_\tau \right|^2 \dif{\tau} + v_z(t,z,f) \sum_{i=1}^d (L^i_s + M^i_s) \right]. \]
This means that the controls are chosen so as to minimize the long-run average deviations of the controlled process $\Xi$ from zero, subject to proportional adjustment costs. Note that in this problem, the state variables $(t,z,f)$ of the original problem are frozen, so that the uncontrolled $\Xi$ is a Brownian motion.

To solve this problem numerically, we approximate the controls $L$, $M$ by absolutely continuous trading rates of the form
\[  L_t^s = \int_0^s \ell^i(\Xi_\tau) \dif{\tau} \quad \text{and} \quad M_s^i = \int_0^s m^i(\Xi_\tau) \dif{\tau}, \]
for functions $\ell$ and $m$ bounded by a finite constant $K$.\footnote{As the artificial constraint $K$ tends to infinity, we then expect to approach the solution of the original problem.} With this restriction, the dynamics of the controlled process are 
\[  \dif{\Xi^i_s} = \nu^i(\Xi;\ell, m) \dif{s} + \alpha \dif{W_s}, \]
where $\nu^i(\xi;\ell,m) = (\ell^i - m^i)(\xi)$. The corresponding dynamic programming equation for the restricted version of~\eqref{eq:ergodic2} in turn is
\[  \min_{\substack{i=1,\dots,d \\ \ell^{i},m^i \in [0, K]}} \left( \mathcal{L}^{\ell,m} w(\xi) + f(\ell, m, \xi) \right) = -a, \quad \forall \xi \in \R^d, \]
where 
\begin{equation}\label{eq:Llm}
\mathcal{L}^{\ell,m} w(\xi) = \nu(\xi; \ell, m)^\top \pd{w}{\xi}(\xi) + \frac{1}{2} \Tr \left[ \alpha \alpha^\top \dmd{w}{2}{(\xi^i)}{}{(\xi^j)}{}(\xi) \right]
\end{equation}
and
\[  f(\ell, m, \xi) = - \frac{1}{2} \left| \sigma_S^\top \xi \right|^2 v_{zz} + v_z \sum_{i=1}^d (\ell^i + m^i). \]
Now, truncate the state space for the control problem to a large finite domain in $\mathbb{R}^d$, and consider a discretization $\mathcal{D} \subset \R^d$ of this set. The approximation $\mathcal{L}^{\ell,m}_\mathcal{D} : \mathcal{D} \mapsto \R$ of the operator $\mathcal{L}^{\ell,m}$ in~\eqref{eq:Llm} outlined in Appendix~\ref{sec:spaceDiscretization} can in turn be interpreted as the transition-rate matrix of a discrete control problem with dynamic programming equation
\[  \min_{\substack{i=1,\dots,d \\ \ell^{i},m^i \in [0, K]}} \left( \sum_{\xi' \in \mathcal{D}} \mathcal{L}^{\ell,m}_\mathcal{D}(\xi,\xi') w(\xi') + f(\ell, m, \xi) \right) = -a, \quad \forall \xi \in \mathcal{D}. \]
If the truncated domain is sufficiently large, the probability of $\Xi$ reaching its boundary is small, so that the corresponding boundary conditions can be chosen arbitrarily, as long as the discretized differential operator can be interpreted as the transition-rate matrix of some discrete control problem. 
The key advantage of this scheme is that the bound $K$ ensures that the transition probabilities are bounded away from 0, enabling us to represent the problem as a continuous time Markov decision process for which standard policy iteration techniques apply. More specifically, this discrete problem can be solved using the following policy iteration algorithm by choosing an initial policy $(\ell_0,m_0)$, e.g., $\ell_0,m_0 \equiv 0$, and then iterating the following steps:
\begin{enumerate}
    \item Compute $(w_j, a_j) \in \R^{|\mathcal{D}|} \times \R^+$ as the solution of
	\[  \left( \sum_{\xi' \in \mathcal{D}} \mathcal{L}^{\ell_j,m_j}_\mathcal{D}(\xi,\xi') w_j(\xi') + f(\ell_j, m_j, \xi) \right) = -a \quad \forall \xi \in \mathcal{D}. \]
	Note that these are $|\mathcal{D}|$ equations for $|\mathcal{D}| + 1$ unknowns. The missing equation is obtained by normalizing $w$ as in Section \ref{sec:homogenization}.
    \item Find solutions $\ell_{j+1}$ and $m_{j+1}$ to the $|\mathcal{D}|$ minimization problems
	\[  \ell_{j+1}(\xi), m_{j+1}(\xi) \in \argmin_{\mathclap{\substack{i=1,\dots,d \\ \ell^{i}, m^i \in [0, K]}}} \,\, \left( \sum_{\xi' \in \mathcal{D}} \mathcal{L}^{\ell,m}_\mathcal{D}(\xi, \xi') w_j(\xi') + f(\ell, m, \xi) \right), \]
	and return to the previous step.
\end{enumerate}
The iteration is terminated when the difference between $a_j$ and $a_{j-1}$ is small enough. It is known that this difference converges to 0 in finite time (cf., e.g., \cite{puterman}). Although this bound is very large for a general policy iteration scheme, it has been observed that policy iterations typically converge very quickly---often in fewer than 20 iterations~\cite{santos.rust.04}. The fast convergence is attained thanks to the scheme's close connections to Newton's method. For more details on these connections, as well as the convergence rate, compare~\cite{puterman.brumelle.79,santos.rust.04,bokanowski.et.al.09}.

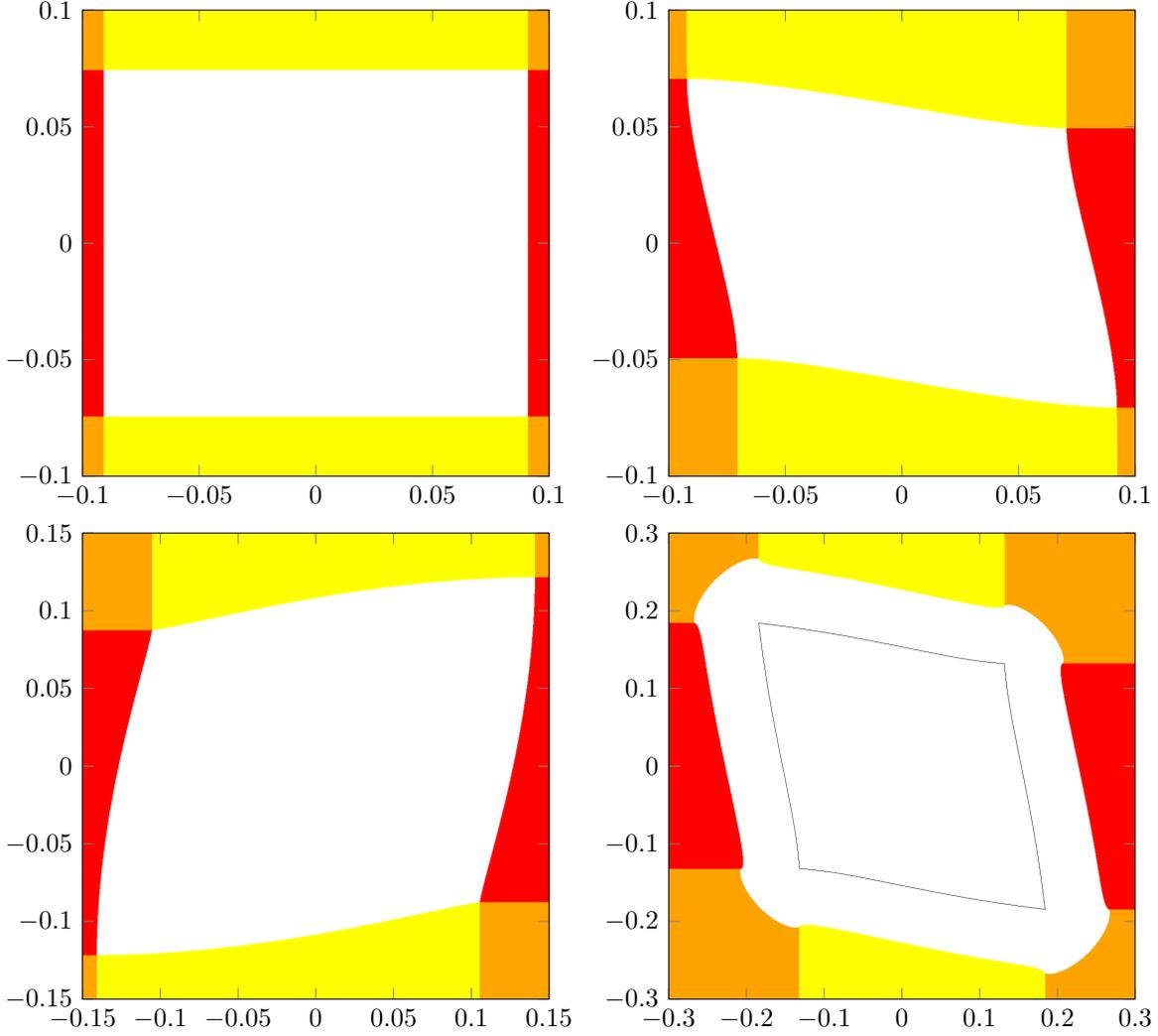
\begin{figure}
    \centering
    \input{Figures/policy_iteration.tex}
    \caption{\label{fig:policy_iteration}Asymptotic no-trade regions for different parameters ($\mu_i$ and $\sigma_i$ are the expected excess return and volatility of risky asset $i=1,2$; $\rho$ is the correlation between their driving Brownian motions). The axes and interpretation of the transaction cost is like in \eqref{eq:ntapproxpower}.}
	\vspace{0.2cm}
	\resizebox{\textwidth}{!}{\begin{tabular}{lcccccccccccr}
		\hline
		rel. pos.    & $\delta$ & $\gamma$ & $r$  & $\mu_1$ & $\mu_2$ & $\sigma_1$ & $\sigma_2$ & $\rho$ & $\lambda_f$ & $\lambda_p$ &    $z$    \\ \hline
		Top left     &    1    &    2     & 0.03 &  0.08   &  0.08   &    0.4     &    0.33     &  0.00  &    0     &     3\%     & \$50'000 \\
		Top right    &    1    &    2     & 0.03 &  0.08   &  0.08   &    0.4     &    0.33     &  0.30  &    0     &     3\%     & \$50'000 \\
		Bottom left  &    1    &    2     & 0.03 &  0.08   &  0.08   &    0.4     &    0.33     &  -0.30  &    0     &     3\%     & \$50'000 \\
		Bottom right &    1    &    3     & 0.03 &  0.08   &  0.08   &    0.4     &    0.4     &  0.30  &    \$1     &     3\%     & \$50'000 \\
		\hline
	\end{tabular}
}
\end{figure}

Solving the $|\mathcal{D}|$ optimization problems in the second step of each iteration may seem daunting at first glance. However, when trading is only conducted through the safe account but not directly between risky assets, the solution of this problem is in fact explicit. Also in more general settings, the optimization problems are entirely independent of each other, so that their solution can be fully parallelized.

For simplicity, the chosen model is a Black--Scholes market consisting of two risky assets, where an agent optimizes the power utility of consumption over an infinite horizon (impatience parameter $\delta$), like in Section \ref{ssec:generalprefs} or \cite{possamai.al.15}. Note that this choice only appears in the above problem through $v_{zz}$, $v_z$, and $\alpha$, and in this case
\[  \alpha = (I_d - \theta_z^\top 1_d) \diag[\theta] \sigma_S, \]
where $I_d$ is the $d$-dimensional identity matrix, $\diag[\theta]$ is the matrix with diagonal $\theta$ and other elements zero, and $1_d=(1,\dots,1)^\top \in \mathbb{R}^d$.

Optimal strategies computed using this algorithm are depicted in Figure~\ref{fig:policy_iteration}. These asymptotic no-trade regions should be interpreted as follows. Each axis represents the deviation of a risky weight from its frictionless target. The white region indicates no trading, whereas the other regions describe trading in the assets. Except in the corners, trading is only performed in one asset at a time, inducing vertical or horizontal movements of the portfolio position. For example, the top left figure has a white region of half-width 0.091, meaning that it is optimal to trade risky asset 2 (only) when its current weight from the frictionless optimizer by 9.1\% of current wealth.

The policy iteration scheme presented here can readily be generalized to more complex models.
For example, details and justification for such generalizations can be found in \cite{altarovici.al.16} for a model with proportional and fixed costs like in Section~ \ref{ss.dpfixed}. The output of the algorithm in this setting is illustrated in the bottom right panel in Figure~\ref{fig:policy_iteration}. There, the solid line inside the no-trade region is the rebalancing target, to which the portfolio is readjusted once the boundaries of the no-trade region are breached.

%% file: Figures/policy_iteration.tex
\begin{tikzpicture}
	\pgfplotsset{every axis/.append style={axis on top,
			width=0.55\textwidth,
			enlargelimits=false,
			axis equal image,
			scaled ticks=false,
			tick label style={/pgf/number format/fixed},
			font=\small,
		}
	}
	\matrix{
		\begin{axis}
			\addplot graphics[xmin=-0.1, ymin=-0.1, xmax=0.1, ymax=0.1] {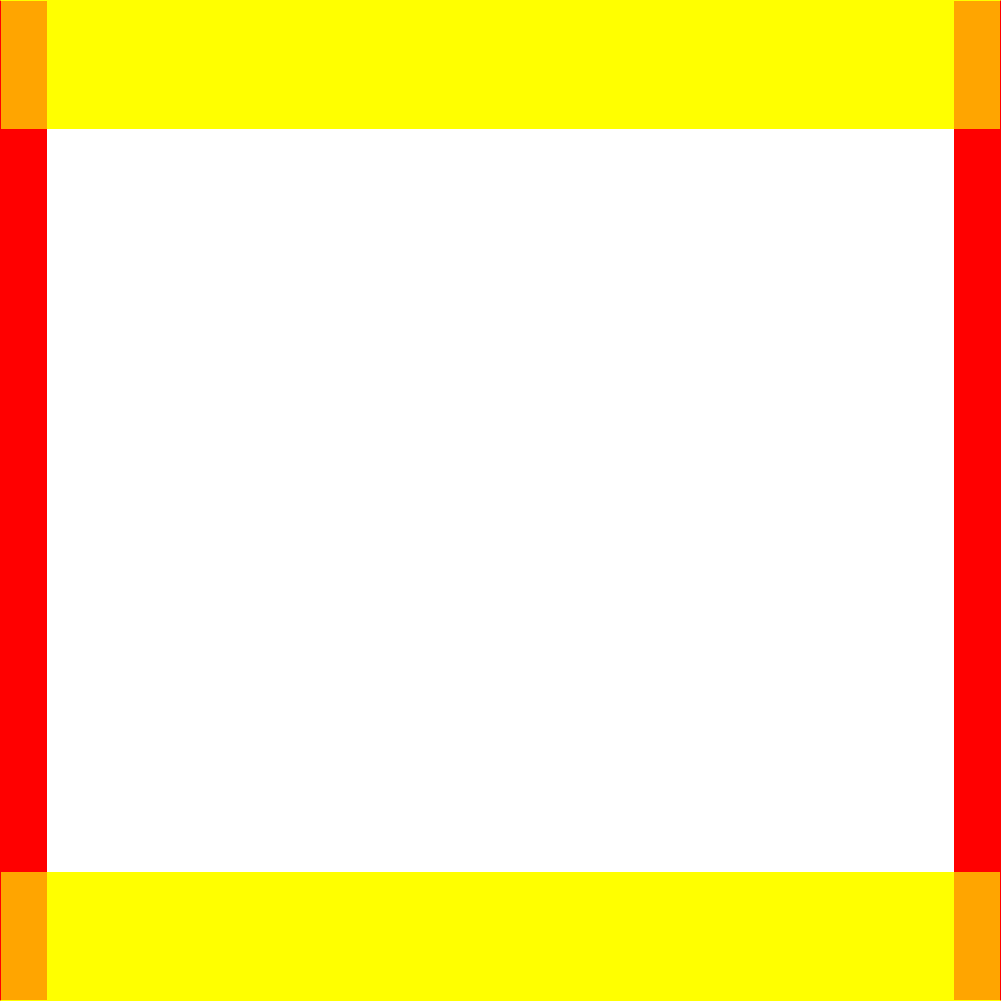};
		\end{axis}
		&
		\begin{axis}
			\addplot graphics[xmin=-0.1, ymin=-0.1, xmax=0.1, ymax=0.1] {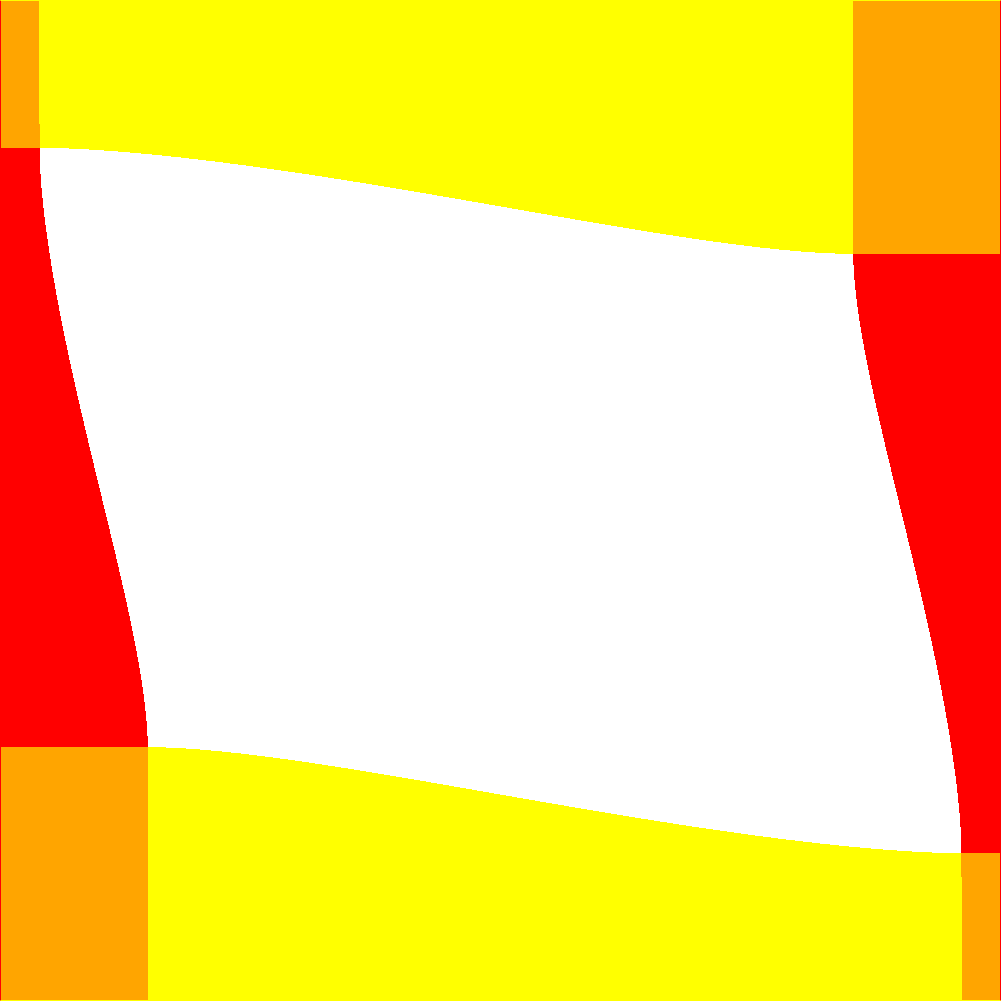};
		\end{axis} \\

		\begin{axis}
			\addplot graphics[xmin=-0.15, ymin=-0.15, xmax=0.15, ymax=0.15] {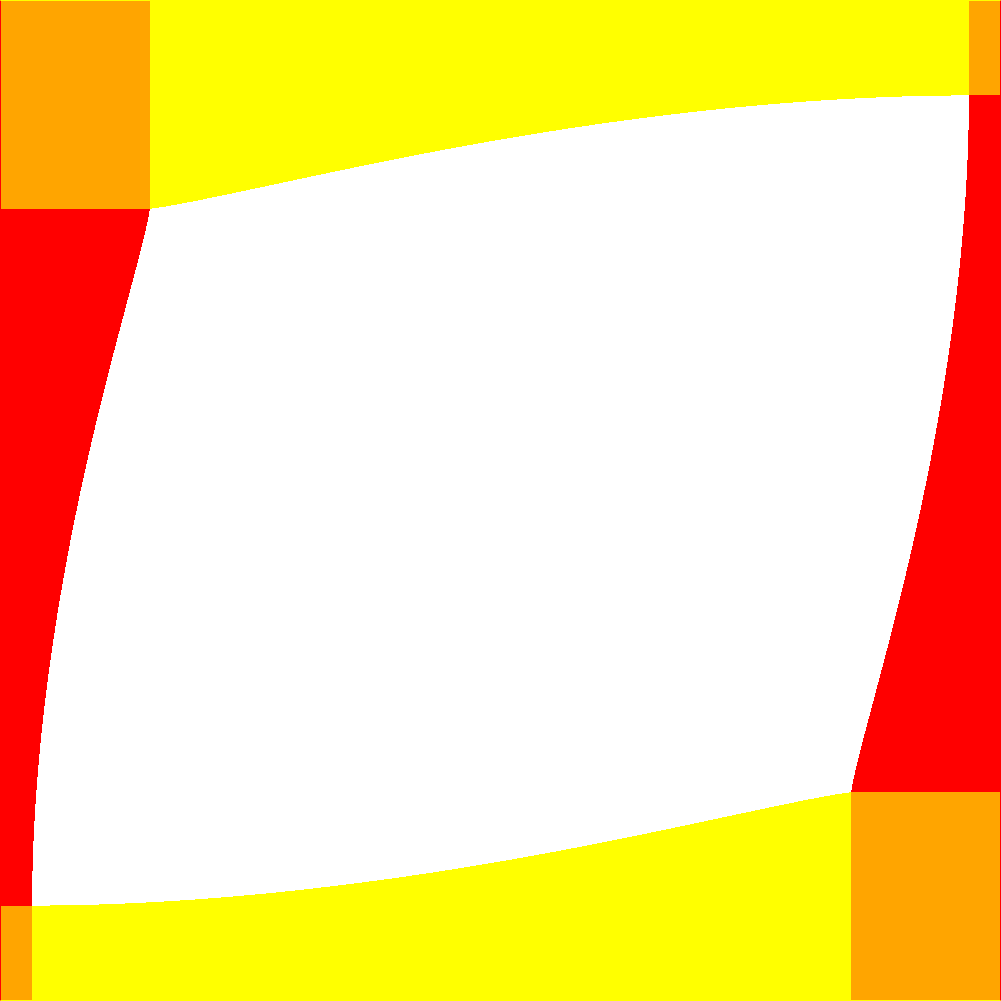};
		\end{axis}
		&
		\begin{axis}
			\addplot graphics[xmin=-0.3, ymin=-0.3, xmax=0.3, ymax=0.3] {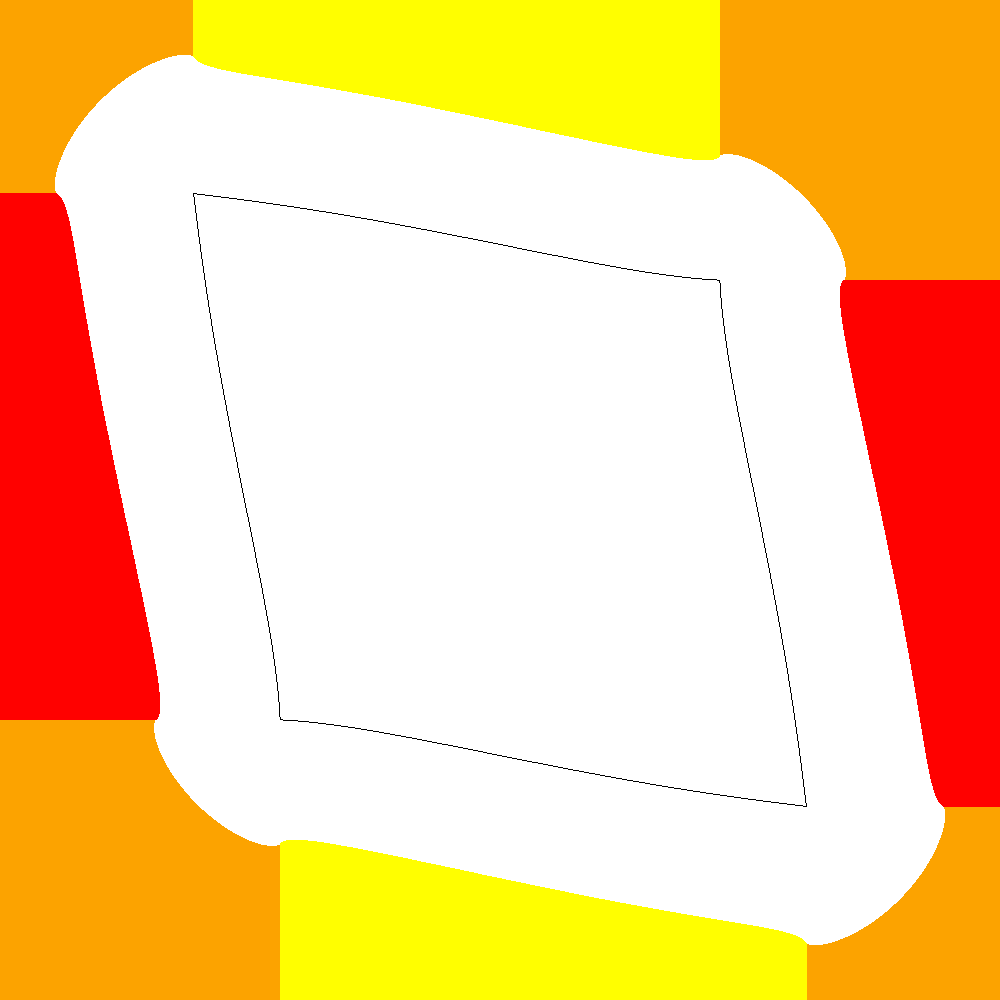};
		\end{axis} \\
	};
\end{tikzpicture}

%% file: appendix.tex
\appendix
\newpage
\section{Kim--Omberg value function}
\label{sec:KOValueFunction}
Consider a power utility function with constant relative risk aversion $\gamma$. As shown in \cite{kim.omberg.96}, the value function for the Kim--Omberg model (\ref{eq:risky}, \ref{eq:oudyn}) then is of the following form:
\[ v(u,z,f) = \frac{z^{1-\gamma}}{1-\gamma} \exp \left(A(u) + B(u)f + \frac{1}{2} C(u) f^2 \right).  \]
Define
\begin{align*}
    b = 2 \left( \frac{1 - \gamma}{\gamma} \frac{\sigma_F}{\sigma_S} \rho - \kappa \right), \quad \eta = \sqrt{b^2 - 4 \frac{1-\gamma}{\gamma} \left(\frac{\sigma_F}{\sigma_S}\right)^2 \left(1 + \frac{1 - \gamma}{\gamma} \rho \right)}.
\end{align*}
In the empirically relevant case when $\gamma>1$ and $\rho<0$, the discriminant $\eta$ is positive, so that $A$, $B$, and $C$ can be identified as the ``normal solution'' of \cite{kim.omberg.96}:
\begin{equation*}
    \begin{aligned}
	C(u) &= \frac{1 - \gamma}{\gamma} \frac{2}{\sigma_S^2} \frac{1 - \exp(-\eta(T-u))}{2 \eta - (b + \eta)\left(1 - \exp\Big(-\eta (T-u)\Big)\right)}, \\
	B(u) &= 4 \frac{1 - \gamma}{\gamma} \frac{\kappa \Fbar}{\sigma_S^2} \frac{1 - \exp( - \eta (T-u)/2)}{2 \eta^2 - \eta (b + \eta)\left(1- \exp\Big(-\eta (T-u)\Big)\right)}, \\
	\text{and} \quad A(u) &= \frac{1 - \gamma}{\gamma} \left( \gamma r + \frac{2 \kappa^2 \Fbar^2}{\sigma_S^2 \eta^2} + \frac{\sigma_F^2}{\sigma_S^2 (\eta - b)} \right) (T-u) \\
	&\quad + \frac{1 - \gamma}{\gamma} \frac{4 \kappa^2 \Fbar^2}{\sigma_S^2} \frac{(2 b + \eta) \exp(-\eta (T-u)) - 4 b \exp(-\eta (T-u)/2) + 2 b - \eta}{\eta^3 (2 \eta - (b + \eta) (1 - \exp(- \eta (T-u))))} \\
	&\quad + \frac{1-\gamma}{\gamma} \frac{2 \sigma_F^2}{\sigma_S^2} \frac{\log\left| 2 \eta - (b + \eta) (1 - \exp(-\eta (T-u)))\right|}{2 \eta (\eta^2 - b^2)}.
    \end{aligned}
\end{equation*}

\newpage
\section{Discretization scheme for policy iteration}
\label{sec:spaceDiscretization}
For the numerical computations in Section \ref{sec:policy_iteration}, the differential operator of the ergodic control problem needs to be discretized. To interpret the discretized operator as the transition rate matrix of some (continuous-time) Markov decision process, the following discretization scheme is used:
\begin{align*}
    \dpd{w}{(\xi^i)}(\xi) & \approx
    \left\{\begin{aligned}
	\frac{w(\xi + e_i h_i) - w(\xi)}{h_i} \quad & \text{if } \nu^i (\xi; \ell,m) > 0, \\
	\frac{w(\xi) - w(\xi - e_i h_i)}{h_i} \quad & \text{if } \nu^i (\xi; \ell) < 0,
    \end{aligned}\right. \\
    \dpd[2]{w}{(\xi_i)}(\xi) & \approx \frac{w(\xi + e_i h_i) - 2 w(\xi) + w(\xi - e_i h_i)}{h_i^2}, \\
    \dmd{w}{2}{(\xi_i)}{}{(\xi_i)}{}(\xi) &\approx
    \begin{cases}
        \begin{aligned}
	    \MoveEqLeft \frac{2 w(\xi) + w(\xi + e_i h_i + e_j h_j) + w(\xi - e_i h_i - e_j h_j)}{2 h_i h_j} \\
	    &- \frac{w(\xi+e_i h_i) + w(\xi - e_i h_i) + w(\xi + e_j h_j) + w(\xi - e_j h_j)}{2 h_i h_j}
        \end{aligned}
	&\quad \text{if } A_{i,j} > 0, \\
        \begin{aligned}
	    \MoveEqLeft - \frac{2 w(\xi) + w(\xi + e_i h_i - e_j h_j) + w(\xi - e_i h_i + e_j h_j)}{2 h_i h_j} \\
	    & + \frac{w(\xi+e_i h_i) + w(\xi - e_i h_i) + w(\xi + e_j h_j) + w(\xi - e_j h_j)}{2 h_i h_j}
        \end{aligned}
	&\quad \text{if } A_{i,j} < 0,
    \end{cases}
\end{align*}
for $i,j=1,\dots,d$, $i \neq j$, and where $h_i$ is the grid size in the $\xi^i$-direction. With $A = \alpha \alpha^\top$, the approximation $\mathcal{L}^\ell_\mathcal{D} : \mathcal{D} \mapsto \R$ of the differential operator $\mathcal{L}^\ell$ is then  given by
\begin{align*}
    \mathcal{L}^\ell_\mathcal{D} (\xi, \xi) & = - \sum_{i=1}^d \Bigg( \frac{A_{i,i}}{h_i^2} - \frac{1}{2} \sum_{\substack{j = 1, \\ j\neq i }}^d \frac{|A_{i,j}|}{h_i h_j} \Bigg) - \sum_{i=1}^d \frac{|\nu^i(\xi; \ell)|}{h_i}, \\
    \mathcal{L}^\ell_\mathcal{D} (\xi, \xi + e_i h_i) & = \frac{1}{2} \Bigg( \frac{A_{i,i}}{h_i^2} - \sum_{\substack{j = 1, \\ j\neq i }}^d \frac{|A_{i,j}|}{h_i h_j} \Bigg) + \sum_{i=1}^d \frac{\max\{0,\nu^i(\xi; \ell)\}}{h_i}, \\
    \mathcal{L}^\ell_\mathcal{D} (\xi, \xi - e_i h_i) & = \frac{1}{2} \Bigg( \frac{A_{i,i}}{h_i^2} - \sum_{\substack{j = 1, \\ j\neq i }}^d \frac{|A_{i,j}|}{h_i h_j} \Bigg) + \sum_{i=1}^d \frac{\max\{0,-\nu^i(\xi; \ell)\}}{h_i}, \\
    \mathcal{L}^\ell_\mathcal{D}(\xi, \xi \pm e_i h_i \pm e_j h_j)  &= \frac{\max\{0, A_{i,j}\}}{2 h_i h_j}, \\
    \mathcal{L}^\ell_\mathcal{D}(\xi, \xi \pm e_i h_i \mp e_j h_j)  &= \frac{\max\{0, -A_{i,j}\}}{2 h_i h_j},
\end{align*}
for $i,j=1,\dots,d$ and $i \neq j$. For a finite domain, we will also need the condition
\[  \sum_{\xi' \in \mathcal{D}} \mathcal{L}^{\ell_j}_\mathcal{D}(\xi, \xi') =0. \]
This ensures that the Markov decision process stays within the domain at all times. As long as the domain is chosen large enough to contain the no-trade region this is not a constraint, since the optimal strategy already ensures that the process does not exit the domain.